\documentclass[runningheads]{llncs}

\usepackage[T1]{fontenc}
\usepackage{graphicx}
\usepackage{hyperref}
\usepackage{amssymb}

\usepackage{amsmath}
\usepackage{mathtools}

\usepackage[createShortEnv]{proof-at-the-end}

\usepackage[capitalise]{cleveref}

\usepackage{booktabs}
\usepackage{csquotes}
\usepackage{ulem}
\usepackage{mdframed}
\usepackage{xspace}
\usepackage{fancyvrb}
\usepackage{adjustbox}
\usepackage{relsize}
\usepackage{comment}

\usepackage{mathpartir}
\usepackage{semantic}
\usepackage{stmaryrd}

\usepackage{xparse}
\usepackage{suffix}

\usepackage{pifont}

\newcommand\doubleplus{+\kern-1.3ex+\kern0.8ex}

\newcommand{\rrule}{\textsc}

\newsavebox{\saveboxedarray}

                                        \reservestyle{\command}{\textsf}

\newcommand{\set}[1]{{\{ #1 \}}}

\makeatletter
\DeclareDocumentCommand{\newmathcommand}{ m O{0} m }{%
  \ifcsname\expandafter\@gobble\string#1\space\endcsname
    \expandafter\expandafter\expandafter\let\expandafter\csname old\string#1\expandafter\endcsname\expandafter=\csname\expandafter\@gobble\string#1\space\endcsname
  \else
    \expandafter\let\csname old\string#1\endcsname=#1
  \fi
  \expandafter\newcommand\csname new\string#1\endcsname[#2]{#3}
  \DeclareRobustCommand#1{%
    \ifmmode
      \expandafter\let\expandafter\next\csname new\string#1\endcsname
    \else
      \expandafter\let\expandafter\next\csname old\string#1\endcsname
    \fi
    \next
  }%
}
\makeatother

\newmathcommand{\o}{\circ}

\newcommand{\CatSet}{\mathbf{Set}}
\newcommand{\CatFam}{\mathbf{Fam}}

\newcommand{\McC}{\mathcal{C}}

\newcommand{\op}{\textsf{op}}

\newcommand{\p}{\mathsf{p}}
\renewcommand{\v}{\mathsf{v}}
\newcommand{\ext}[2]{\langle #1, #2 \rangle}
\newcommand{\cext}{\triangleright}

\newcommand{\Ty}[1]{\mathsf{Ty}({#1})}
 \newcommand\TyG{\Ty{\Gamma}}
 \WithSuffix\newcommand\Ty*[1]{\mathsf{\seq{Ty}}_{#1}}
 \WithSuffix\newcommand\TyG*{\Ty*{\Gamma}}
 \newcommand{\Tm}[1]{\mathsf{Tm}({#1})}
 \newcommand{\TmGT}{\Tm{\Gamma}{T}}
 \WithSuffix\newcommand\Tm*[2]{\mathsf{\seq{Tm}}_{#1,#2}}
 \WithSuffix\newcommand\TmGT*{\Tm*{\Gamma}{\seq{T}}}

\newcommand{\bN}{\mathbb{N}}
\newcommand{\bZero}{\mathbb{0}}
\newcommand{\bOne}{\mathbb{1}}

    \newcommand\blackoline[1]{\colorlet{temp}{.}\color{black}\overline{\color{temp}#1}\color{temp}}
    \newcommand{\seq}[1]{\blackoline{#1}}

\newcommand{\bnfalt}{\mathbf{\,\,\mid\,\,}}
\newcommand{\bnfdef}{\mathbf{\ \Coloneqq\ }}
\newcommand{\bnfadd}{{\mathbf{\ +\!\!\Coloneqq\ }}}
\newcommand{\bnfsub}{{\bf - : : =}}

\newcommand{\ie}{i.e.\ }

\newcommand{\ala}{\`a la\xspace}

\definecolor{lightgray}{gray}{0.90}

\newcommand{\mynote}[3][red]
    {{\color{#1} \fbox{\bfseries\sffamily\scriptsize#2}
    {\small$\blacktriangleright$\textsf{\emph{#3}}$\blacktriangleleft$}}~}

\newcommand{\je}[1]{\mynote{JE}{#1}}

\newcommand{\typeType}[1]{\mathsf{Type}_{#1}}

\mathlig{||}{\bnfalt}
\mathlig{::=}{\bnfdef}
\mathlig{+::=}{\bnfadd}
\mathlig{-::=}{\bnfsub}
\mathlig{|>}{\vartriangleright}
\mathlig{<|}{\vartriangleleft}
\mathlig{<=}{\Leftarrow}
\mathlig{=>}{\Rightarrow}
\mathlig{>->}{\rightarrowtail}
\mathlig{|->}{\mapsto}
\mathlig{|=>}{\Mapsto}
\mathlig{==}{=}
\mathlig{**}{\times}
\mathlig{<<}{\langle}
\mathlig{>>}{\rangle}
\mathlig{|=}{\vDash}

\usepackage{wrapfig}

\newcommand{\G}{\Gamma}
\newcommand{\D}{\Delta}
\newcommand{\blank}{\_}

\newcommand{\env}[1]{\texttt{#1}}

\DeclareUrlCommand\DOI{}
\newcommand{\doi}[1]{\href{http://dx.doi.org/#1}{\DOI{doi:#1}}}



\newcommand{\jform}[1]{\fbox{#1}\hspace{\fill}\\}
\newcommand{\jformnb}[1]{\fbox{#1}\hspace{\fill}}

\usepackage{bbold}

\usepackage{scalerel}
\usepackage{float}

\usepackage[capitalise]{cleveref}
\Crefname{theorem}{Thm.}{Thms.}%
\Crefname{lemma}{Lem.}{Lems.}%
\Crefname{corollary}{Cor.}{Corss.}%

\newcommand{\oblset}[1]{\textsc{#1}}

\let\oldAE\AE
\let\oldae\ae
\renewcommand{\AE}{\textsf{\oldAE}}
\renewcommand{\ae}{\textsf{\oldae}}

\usepackage{newunicodechar}
\newunicodechar{⨟}{ \ensuremath{{ \fatsemi }}}
\newunicodechar{⋁}{ \ensuremath{{ \bigvee }}}
\newunicodechar{⊔}{ \ensuremath{\sqcup}}

\newcommand{\match}[4]{\mathtt{case}\ (#1 : #2)\ \mathtt{to}\ #3\ \mathtt{of}\ \{#4\}}

\usepackage{suffix}

 \newcommand{\covers}{\text{\reflectbox{$\sphericalangle$}}}
 \DeclareMathOperator{\envext}{\fatsemi}

 \DeclareMathOperator{\fish}{{<}\!{>}\!{<}}
 \DeclareMathOperator{\chips}{{<}\!{>}\!\!{>}}

 \mathlig{|-}{\vdash}
 \mathlig{<><}{\fish}
 \mathlig{<>>}{\chips}
 \mathlig{++}{\doubleplus}
 \mathlig{[[}{\llbracket}
 \mathlig{]]}{\rrbracket}

 \newcommand{\applySub}[2]{#1\{#2\}}
 \newcommand{\sub}[1]{\{#1\}}

\newcommand{\branch}{=>}

\newcommand{\ctx}{\ \mathsf{ctx}}
\newcommand{\refl}[1]{{\mathtt{refl}_{#1}}}

\newcommand{\seqi}[2]{{{\overrightarrow{#2}}^{#1}}}
\newcommand{\setj}[2]{{\{#2\}_{#1}}}
\newcommand{\singleton}[1]{{\{#1\}}}

\DeclarePairedDelimiter\mpars{\lparen}{\rparen}

\newcommand{\lang}{\oblset{CoverTT}\xspace}

\renewcommand{\env}[1]{\mathbf{#1}}

\newcommand{\Hom}[3][]{\mathsf{Hom}_{#1}\mpars*{#2,#3}}

\newcommand{\inj}{\mathsf{inj}}

\renewcommand{\typeType}[1][{}]{\mathcal{U}_{#1}}

\DeclareFontFamily{U}{dmjhira}{}
\DeclareFontShape{U}{dmjhira}{m}{n}{ <-> dmjhira }{}

\DeclareRobustCommand{\yo}{%
  \mathbf{y}%
}

\usepackage{tikz-cd}
\makeatletter
\def\fixtikzforbreqn#1#2{%
  \protected\edef#1{\noexpand\ifmmode\mathchar\the\mathcode`#2 \noexpand\else#2\noexpand\fi}%
}
\fixtikzforbreqn\tikz@nonactivesemicolon;
\fixtikzforbreqn\tikz@nonactivecolon:
\fixtikzforbreqn\tikz@nonactivebar|
\fixtikzforbreqn\tikz@nonactiveexlmark!
\makeatother

\usepackage{suffix}
\newcommand\pair[2]{\left(#1,#2\right)}
\WithSuffix\newcommand\pair-[2]{(#1,#2)}
\newcommand\triple[3]{\left(#1,#2,#3\right)}
\WithSuffix\newcommand\triple-[3]{(#1,#2,#3)}
\newcommand\family[2][]{\left(#2\right)_{#1}}
\WithSuffix\newcommand\family-[2][]{(#2)_{#1}}
\newcommand\carrier[1]{\underline{#1}}
\newcommand\ordinary[1]{#1_{\mathrm{o}}}
\newcommand\from{\leftarrow}
\newcommand\compose{\mathbin{\circ}}
\newcommand\id[1][]{\mathrm{id}_{#1}}
\newcommand\xto[2][]{\xrightarrow[#1]{#2}}

\newcommand{\emptyCtx}{\cdot}
 \renewcommand{\Tm}[2][]{\mathsf{Tm}_{#1}({#2})}

\newcommand\concept[1]{\emph{#1}}
\newcommand\stress[1]{\emph{#1}}
\renewcommand\concept[1]{\textbf{#1}}
\renewcommand\stress[1]{#1}

\renewcommand\ext[3][]{\langle#2, #3#1\rangle}

\newcommand\lean{\textsc{lean}}
\newcommand{\leanCheck}{\mathsf{({\checkmark}\lean)}}

\newcommand{\scott}[1]{\llbracket #1 \rrbracket}

\AtEndPreamble{%
}
\newcommand\isomorphic{\cong}

\pgfkeys{/prAtEnd/apxproof/.style={
 no link to proof,restate 
  }
}

\pgfkeys{/prAtEnd/apxlem/.style={
all end
  }
}

\pgfkeys{/prAtEnd/global custom defaults/.style={
}
}

\usepackage{color}

\usepackage[numbers]{natbib}

\begin{document}

\title{Coverage Semantics for Dependent Pattern Matching}
\titlerunning{Coverage Semantics for Dependent Pattern Matching}


\author{Joseph Eremondi\inst{1}\orcidID{0000-0002-9631-4826} \and
Ohad Kammar\inst{2}\orcidID{0000-0002-2071-0929}}

\institute{
  Department of Computer Science, University of Regina, Regina, SK, Canada\\
  \email{jeremondi@uregina.ca}
  \and
  {Laboratory for the Foundations of Computer Science, University of Edinburgh, Edinburgh, Scotland, United Kingdom \\
  \email{ohad.kammar@ed.ac.uk}}
}

\maketitle

\begin{abstract}

Dependent pattern matching is a key feature in dependently typed programming.
However, there is a theory-practice disconnect:
while many proof assistants implement pattern matching as primitive,
theoretical presentations give semantics to pattern matching
by elaborating to eliminators. Though theoretically convenient,
eliminators can be awkward and verbose, particularly for complex combinations of patterns.

This work aims to bridge the theory-practice gap by presenting
a direct categorical semantics for pattern matching,
which does not elaborate to eliminators.
This is achieved using sheaf theory
to describe when sets of arrows (terms) can be amalgamated into
a single arrow.
We present a language with top-level dependent pattern matching, without specifying
which sets of patterns are considered covering for a match.
Then,  we give a sufficient criterion for which pattern-sets
admit a sound model: patterns should be in the canonical coverage
for the category of contexts.
Finally, we use sheaf-theoretic saturation conditions to
devise some allowable sets of patterns.
We are able to express and exceed the status quo, giving semantics
for datatype constructors, nested patterns, absurd patterns,
propositional equality, and dot patterns.

\end{abstract}


\section{Introduction}





%
%
Pattern matching is a core feature in dependently typed programming.
With pattern matching one can specify a function consuming an input
by giving  functions for every possible way
that input might have been constructed.
For dependent types, the defined function can be a universally quantified proof,
giving a Curry-Howard analogue of proof-by-cases.

However, there is a disconnect between the theory and practice of dependent pattern matching.
Many dependently typed languages take pattern matching as a built-in user-facing construct:
Coq~\citep{bertotInteractiveTheoremProving2004}, Agda~\citep{norellDependentlyTypedProgramming2009}, and Idris~\citep{bradyIdrisQuantitativeType2021}
all contain a form of dependent pattern matching in their core calculi.
However, most theoretical treatments of dependent types
deal with \textit{eliminators}~\citep{mcbrideEliminationMotive2002}: primitive recursors
with result types dependent on a value
of the eliminated type.
Eliminators and pattern matching are equally expressive, with or without
with Axiom K~\citep{goguenEliminatingDependentPattern2006,cockxPatternMatching2014}.
%

While it is possible to express all pattern matches using eliminators,
it is not always convenient. Some pattern matching features require lengthy translations
when converting to eliminators, such as overlapping patterns, catch-all branches, or matching on
multiple values at once.
Moreover, languages differ in which pattern matches they allow, so every variant of pattern matching requires a new eliminator-translation
to prove consistency.
Even the implementations of dependently typed languages are restricted by eliminators,
since most pattern matches are elaborated into {case trees} with
a 1:1 correspondence between branches and constructors of an inductive type.

%
The contribution of this paper is to narrow the theory-practice divide
with a highly general {syntax} (\cref{sec:syntax}) and {categorical semantics for dependent pattern matching}
(\cref{sec:cwf}).
The semantics is {direct} and {generic}:
pattern matches are translated
directly into semantic objects without desugaring
to eliminators or case trees,
and the semantics is parameterized over an abstract {coverage}
specifying which sets of patterns one can match against.
We investigate the vision set forth by Epigram~\citep{mcbrideEpigramPracticalProgramming2005,mcbrideViewLeft2004} to enable diverse pattern-matching abstractions that go beyond the list of constructors declared by a datatype.
We focus on non-overlapping patterns, but give a potential road map
to supporting overlap.

In constructing our generic semantics, we present
a {general sufficient criterion for when a coverage leads to well-defined pattern matches}
without compromising logical consistency (\cref{sec:sheaf}),
mechanized in Lean~\citep{joey_eremondi_2025_14768609}.
We define this criterion, drawing on parallels between dependent pattern matching
and the theory of sheaves on a site, discovering
that it is sufficient for each allowed set of patterns to correspond
to a cover in the canonical coverage for the semantic category.
Moreover, we use elementary results from sheaf theory to describe a group of {closure operations}
which preserve the canonicity of a coverage (\cref{sec:coverages}). These
give a simple and direct way to model common features like
multi-value matches, nested patterns, and matching on propositional equality proofs.
With the exception of recursion, we achieve feature parity with
the original presentation of dependent pattern matching by  \citet{coquandPatternMatchingDependent1992}.
We conclude with an illustrative example (\cref{sec:example}) and
related and future work (\cref{sec:discussion}).

A major contribution of our work is expressing pattern matching in the language
of categories and sheaves.
Our approach is semantic: instead of elaborating pattern matching syntax,
we take pattern matches as  primitive
and work directly in the semantic domain,
avoiding the need to consider eliminators as a syntactic primitive.
The connection has been implicit for many decades, but we make it formal.
That said, we only assume basic knowledge of functors, pullbacks, and slice categories,
and we present all the required sheaf theory.

\section{\lang: The Source Language}
\label{sec:syntax}

We begin with a variant of Martin L\"of Type Theory, called \lang,
parameterized over
 which sets of patterns  can be matched against. 
 Drawing from sheaf-theory terminology, when a set of patterns is permissible
 on the left-hand side of a pattern match we call it a \concept{cover},
 and say it is \concept{covering}.
The set of all covers together is called a \concept{coverage}.
The main distinct feature of \lang
 is that it is parameterized over
a coverage.

\subsection{The Anatomy of a Datatype}
\label{subsec:anatomy}

As they are presented in most dependently typed languages, inductively defined
datatypes conflate four different concepts.
Consider the quintessential inductive family of length-indexed vectors:
\begin{flalign*}
  &\mathtt{data}\ \mathsf{Vec}\ (A : \typeType{}) : (m : \bN) -> \typeType\ \mathtt{where}\\
  &\qquad \mathsf{nil} : \mathsf{Vec}\ A\ 0\\
  &\qquad \mathsf{cons} : (n : \bN) -> A -> \mathsf{Vec}\ A\ n -> \mathsf{Vec}\ A\ (n+1)
\end{flalign*}
This definition implicitly relies on the following concepts, the separation of which motivates the  design \lang
(as exhibited in \cref{subsec:vec-example}).
\begin{itemize}
  \item \textbf{Coproducts:}  An inductive type behaves like the sum of its constructor types. For $\mathsf{Vec}$,
        it behaves like $\bOne + (A \times \mathsf{Vec}\ A\ n)$.
        In \lang, each inductive $I$ is declared at the top level to have a finite collection
        of constructors
        $D^{I}_{1}\ldots D^{I}_{n}$, each of which has a type ending in $I$.
  \item \textbf{Dependent fields:} each constructor has a dependent product type, so the
        arguments to a constructor are a curried dependent record, where the types
        of later fields can depend on the values of earlier ones.
        For $\mathsf{Vec}$, in $\mathsf{cons}$ the return type and the
        type of the tail of the list depend
        on the earlier parameter $n$.
        In \lang, the type of a constructor is given as a telescope,
        where the types of later entries are allowed to depend on the values of previous entries.
  \item \textbf{Indexing:} each constructor implies a specific equation about the index values.
        For $\mathsf{Vec}$, $\mathsf{nil}$ restricts that $m = 0$ and $\mathsf{cons}$ restricts
        that $m = n+1$.
        In \lang,
     we use
        \concept{Fording}~\citep{mcbrideDependentlyTypedFunctional2000}, where each constructor has an identical return type,
        but may constrain type parameters using equality-proof fields.
        We treat equality as primitive and make it the only way to constrain indices of a constructor,
        simplifying the models of \lang.
  \item \textbf{Recursion:} inductive types can refer to themselves in fields, except to the left of a function arrow, a condition known as strict positivity.
        Typically, one can define structurally recursive functions over $\mathsf{Vec}$.
        Here, $\mathsf{Vec}$ occurs as a field type for $\mathsf{cons}$, which is allowed
        because it is not to the left of an arrow type.
        We omit recursion from \lang, as we believe it requires a separate toolkit of abstractions.
        In \cref{subsec:future-rec} we discuss its possible addition. In examples,
        we refer to some inductive types defined using self-reference, such as vectors or natural numbers.
\end{itemize}

\paragraph{Restricting to Top-Level Datatypes}

\lang only allows
data types and pattern matches to be declared at the top level.
The parameter and constructor types of datatypes
must be typeable in the empty context, though they can refer to other data types.
Likewise, the branches and motive of a pattern match must be typeable in the empty context.
These assumptions simplify the presentation of \lang while reflecting how
datatypes are implemented in languages like Agda, where
declarations in a non-empty context
are desugared into top level declarations with extra parameters.
Treating nested patterns and arbitrary coverages is an interesting problem
even with top-level matches, and requires substantial technical developments even to handle ordinary matching.

\subsection{Syntax and Typing}

\begin{figure*}[!b]
\begin{flalign*}
   \textsc{Term} \ni s,S,t,T ::= & \quad
    x
    \bnfalt \typeType
    \bnfalt \Pi(x:S)\ldotp T
    \bnfalt \lambda x \ldotp t
  \bnfalt t\ s
    \bnfalt s =_{T} t
  \bnfalt \refl{t}\\&\qquad
    \bnfalt I\ \env{t}
    \bnfalt D^{I}\ \env{t}
  \bnfalt \match{\env{t}}{\Gamma}{T}{\seqi{i}{\Delta_{i}\ldotp\env{s}_{i} \branch t}} \\
  \textsc{Ctx} \ni \Gamma, \Delta, \Xi ::= & \quad  \cdot \bnfalt \Gamma, (x : T) \\
  \textsc{Subst} \ni \env{s}, \env{t} ::=  & \quad \cdot \bnfalt \env{s} \envext t
\end{flalign*}
  \centering
    \fbox{$\Gamma |- t : T$ \textit{(Well Typed Terms/Types)}}
    \quad \fbox{$\setj{i}{t_{i}} \covers \Xi$ \textit{ (Covering Patterns) }}
    \begin{mathpar}
      \inferrule[TypeConv]{\Gamma |- t : S \\\\ \Gamma |- S \equiv T : \typeType}{\Gamma |- t : T }

      \inferrule[TyVar]{|- \Gamma \ctx \\\\ (x : T) \in \Gamma}{\Gamma |- x : T}

      \inferrule[TyPi]{\Gamma |- S : \typeType \\\\
        \Gamma, (x : S) |-  T : \typeType
      }{
        \Gamma |- \Pi(x : S)\ldotp T : \typeType
      }

      \inferrule[TyApp]{\Gamma |- t : \Pi(x : S)\ldotp T\\\\
        \Gamma |- s : S
      }{
        \Gamma |- t\ s : [s/x]T
      }

      \inferrule[TyLam]{\Gamma, (x : S) |- t : T}{
        \Gamma |- \lambda x \ldotp t : \Pi(x : S)\ldotp T
      }

      \inferrule[TyRefl]{\Gamma |- t : T}{
        \Gamma |- \refl{t} : t =_T t
      }

      \inferrule[TyInd]{
        \mathsf{Params}(I) := \Delta\\
        \Gamma |- \env{t} => \Delta
      }{\Gamma |- I\ \env{t} : \typeType}

      \inferrule[TyEq]{\Gamma |- T : \typeType\\
        \Gamma |- s : T\\\\
        \Gamma |- t : T
      }{
        \Gamma |- s =_T t : \typeType
      }

      \inferrule[TyCtor]{
        \mathsf{Fields}(D^I) := \Delta, \Xi\\
        \Gamma |- \env{s} => \Delta\\\\
        \Gamma |- \env{t} => [\env{s}/\D]\Xi
      }{\Gamma |- D^I\ \env{t} : I\ \env{s}}

    \inferrule[TyCase]{
      |- \Xi \ctx\\
      \Gamma |- \env{t_{scrut}} => \Xi\\
      \Xi |- T_{motive} : \typeType\\
      \seqi{i}{ |- \Delta_i \ctx  }\\      \seqi{i}{\Delta_i |- \env{s}_{i} => \Xi}\\\\
      |- \setj{i}{\env{s_{i}}} \covers \Xi\\
      \seqi{i}{\Delta_i |- t_i : [\env{s}_{i}/\Xi]T_{motive}}
    }{
      \Gamma |- \match{\env{t_{scrut}}}{\Xi}{T_{motive}}{\seqi{i}{ \Delta_i\ldotp \env{s}_{i} \branch t_{i}}}
      : [\env{t_{scrut}} / \Xi]T_{motive}
        }
    \end{mathpar}
  \caption{\lang: Term Typing}
  \label{fig:typing}
\end{figure*}

The syntax for \lang below is standard, except for the
separation of concerns from \cref{subsec:anatomy}.
\Cref{fig:typing} gives the syntax, along with typing rules for \lang.

    Overline arrows denote sequences, while
    bold metavariables denote dependent sequences, e.g. substitutions. Variables are assigned types from the context.
    Typing for dependent functions and equality is standard,
    with rules for their types, introduction, and elimination.
    The equality type is intensional, and there is no reflection rule
    by which propositional equalities can be made judgmental equalities.
    Nevertheless, \lang is consistent with
    models that identify all propositionally-equal terms.
    We have no J-axiom, instead using the coverage to specify how to
    match on $\refl{t}$.
    We have one universe which does not have a type, leaving a universe hierarchy for future work.

\subsubsection{Contexts and Substitutions}

    \Cref{fig:ctx-typing} also specifies well-formedness rules for contexts and substitutions.
    Contexts are sequences of typed variables, where later types may refer to variables from earlier in the context.
%
Substitutions are the inhabitants of contexts. The rules are like an iterated version
of dependent pairs:
the type of the later values in the substitutions
may depend on the earlier values.
We borrow the notation $\Gamma |- \env{t} => \Delta$ from \citet{hofmannSyntaxSemanticsDependent1997}, since such a substitution
corresponds to a morphism $\Gamma -->  \Delta$ in the models we define.
We use the notation $[\env{s}/\Delta]t$ to denote the simultaneous substitution of
the variables bound in $\Delta$
by the terms of $\env{s}$ in $t$. If
$\Gamma |- \env{s} => \Delta$ and
$\Delta |- t : T$,
then $\Gamma |- [\env{s}/\Delta] t : [\env{s}/\Delta]T$.

\subsubsection{Datatype and Pattern Matching Syntax}

We assume a fixed collection of inductive type constructors,
along with data constructors.
For each datatype there is a fixed context of parameters $\mathsf{Params}(I)$
and, for each constructor, fields $\mathsf{Fields}(D^{I})$,
such that $|- \mathsf{Params}(I), \mathsf{Fields}(D^{I})$, i.e., both are well-formed,
and the fields may depend on the parameters.
The type $I\ \env{t}$ denotes the type constructor $I$ applied to parameters $\env{t}$.
$D^{I}\ \env{t}$ is the data constructor $D$ for the type $I$, given $\env{t}$
for its fields. We omit $I$ in $D^{I}$ when it is clear from context.

The pattern matching form is a nameless version of defining functions by multiple pattern
matching clauses, as in Agda or Idris.
The term
\begin{displaymath}
\match{\env{t}}{\Xi}{T}{\seqi{i}{\Delta_{i}\ldotp \env{s}_{i} \branch t_{i}}}
\end{displaymath}
denotes a match on \concept{scrutinees} $\env{t}$
of context-type $\Xi$, producing a result of the \concept {motive} type $T$, where $T$ may refer to the variables
bound in $\Xi$.
The branches, indexed by $i$,
each have a left--side pattern $\env{s_{i}}$, which may contain pattern variables
from the context $\Delta_{i}$, and which produces a result $t_{i}$.
The scrutinee is a substitution  because pattern matching functions
can take multiple arguments, and the types of later arguments can depend on the values
of earlier ones.

\begin{figure*}[t]
      \centering

      \fbox{$|- \Gamma \ctx$ \textit{ (Well Formed Contexts) }}
      \qquad
  \fbox{$\Gamma |- \env{t} => \Delta$ \textit{ (Well Formed Substitutions)}}
  \begin{mathpar}
    \inferrule[CtxNil]{  }{|- \cdot \ctx}

    \inferrule[CtxCons]{
      |- \Gamma \ctx \\
      \Gamma |- T : \typeType\\
    }
    {|- \Gamma, (x : T) \ctx}

    \inferrule[EnvNil]{  }{\Gamma |- \cdot => \cdot}

    \inferrule[EnvCons]{
      \Gamma |- \env{s} => \Delta \\
      \Gamma |- t : [\env{s}/\Delta]T}
    {\Gamma |- \env{s} \envext t => \Delta, (x : T) }
  \end{mathpar}





      \caption{Typing: Contexts and Substitutions}
      \label{fig:ctx-typing}
    \end{figure*}

\subsubsection{Typing Pattern Matches}

The \rrule{TyCase} typing rule (\cref{fig:typing}) is the  most important rule.
To type
\begin{math}
  \Gamma |- \match{\env{t_{scrut}}}{\Xi}{T_{motive}}{\seqi{i}{ \Delta_{i}\ldotp \env{s}_{i} \branch t_{i}}}
\end{math},
the scrutinee inhabits some $\Xi$.
Because pattern matching is \stress{dependent}, the motive
$T_{motive}$ is indexed over the scrutinee type.
The pattern $\env{s_{i}}$ for each branch inhabit the scrutinee type $\Xi$ in the context of its pattern variables $\Delta_{i}$.
Each branch is typed against the context of its pattern variables,
and inhabits the motive for the scrutinee value given by that branch's pattern,
e.g., $\env{s_{i}}$.
Finally, the entire match inhabits the motive type,
instantiated to the scrutinee.
Critically, the motive $T_{motive}$ and each branch $t_{i}$ must be typeable in the closed context $\Delta_{i}$, making no reference to $\Gamma$.
This restriction matches practice: in Agda and Idris,
pattern matching elaborates to top level declarations.
We give an example in \cref{subsec:vec-example}.

\subsubsection{The Generic Coverage Relation}

In a match, the patterns $\seq{\env{s_{i}}}$ must be covering.
At no point do we require the patterns of a match to correspond to
constructors of an inductive type, or even that the matched upon type be inductive.
Instead, we appeal to an arbitrary judgment $\setj{i}{\env{s_{i}}} \covers \Xi$,
relating sets of substitutions to contexts.
This relation
is to be read as ``the patterns $\env{s_{1}} \ldots \env{s_{n}}$ are a total decomposition of the context $\Xi$''.
This relation replaces the usual condition that there must be a case for each constructor
of the datatype. It
is the parameter by which we tune \lang,
and can take many forms, from requiring a single scrutinee with exactly one branch per constructor, to allowing multiple scrutinees with arbitrary nested patterns and absurd branches omitted.
\Cref{subsec:sheaf-pat}
explores the conditions a coverage must satisfy to result in a well-behaved type theory.
In general, we expect that a coverage will consist of a basis set of coverings, containing at least variables and constructors for each datatype, which can be closed under
composition, concatenation, etc.
For generality, we do not require closure conditions, but we show
in \cref{sec:coverages} they are always permissible.


\subsubsection{Computational Rules}

\Cref{fig:defeq} gives definitional equality for \lang,
omitting structural (reflexivity, symmetry, transitivity) and congruence rules.
Rule \rrule{EqApp} is the usual $\beta$-reduction.
In \rrule{EqMatch}, we reduce a match when the scrutinee is
 $[\env{t_{mat}}/\Delta_j]\env{s}_j$ for some
    $\Gamma |- \env{t_{mat}} => \Delta_{j}$,
    i.e., it is a pattern applied to values for the pattern variables of $\env{s_{j}}$.
   This substitution $\env{t_{mat}}$ instantiates the pattern variables
   in $t_{j}$, which is the value of the entire match.

\begin{figure*}
  \centering
  \jformnb{$\Gamma |- t = s : T$ \textit{(Term Definitional Equality)}}~\jform{$|- \Gamma \equiv \D\ \mathsf{ctx}$ \textit{ (Equal Contexts) }}~~~~\jform{$\Gamma |- \env{t} => \Delta$ \textit{ (Equal Substitutions)}}
  \begin{mathpar}
    \inferrule[EqMatch]{
      |- \Xi \ctx\\
      \seqi{i}{ |- \Delta_i \ctx  }\\
      j \in \set{1 \ldots i}\\
      \Gamma |- \env{t_{mat}} => \Delta_j\\
      \Xi |- T : \typeType\\
            \seqi{i}{\Delta_i |- \env{s}_{i} {=>} \Xi}\\
      |- \setj{i}{\env{s}_{i}} \covers \Xi\\
      \seqi{i}{\Delta_i |- t_{i}  : [\env{s}_{i}/\Xi]T}
    }{
      \Gamma |- \match{[\env{t_{mat}}/\Delta_i]\env{s}_j}{\Xi}{T}{\seqi{i}{ \Delta\ldotp \env{s}_{i} \branch t_{i}}}
        \equiv [\env{t_{mat}}/\Delta_j]t_{j} : [\env{t_{mat}} / \D_j][\env{s}_{j}/\Xi]T
        }

      \inferrule[EqApp]{\Gamma, (x : S) |- t : \Pi(x : S)\ldotp T\\
        \Gamma |- s : S
      }{
        \Gamma |- (\lambda x \ldotp t)\ s \equiv [s/x]t : [s/x]T
      }

      \inferrule[EqCtxNil]{  }{|- \cdot \equiv \cdot \ctx}

      \inferrule[SubNil]{  }{\Gamma |- \cdot \equiv \cdot : \cdot}

    \inferrule[EqCtxCons]{
      |- \Gamma \equiv \Gamma' \ctx \\
      \Gamma |- T \equiv T' : \typeType\\
    }
    {|- \Gamma, (x : T) \equiv \Gamma', (x : T') \ctx}

    \inferrule[SubCons]{
      \Gamma |- \env{s} \equiv \env{s'} : \Delta \\
      \Gamma |- t \equiv t' : [\env{s}/\Delta]T}
    {\Gamma |- \env{s} \envext t \equiv \env{s'} \envext t' : \Delta, (x : T) }
  \end{mathpar}
  \caption{Definitional Equality: Computational Rules}
  \label{fig:defeq}
\end{figure*}
%



   \subsection{Example: Vectors}
   \label{subsec:vec-example}

To see these constructs concretely, we show how length-indexed vectors from \cref{subsec:anatomy} would be represented in our system,
along with a type safe head function.
In \lang, vectors can be defined using labelled sums. Notice the equality-type fields which
encode the index constraints via Fording.
\begin{flalign*}
  & \mathsf{Params}({\mathsf{Vec}})
     := \cdot,
     (A : \typeType),(n : \bN)&\\
  &\mathsf{Fields}(\mathsf{Nil}^{\mathsf{Vec}}) :=
      \cdot,
     (A : \typeType),(n : \bN), (eq : n = 0)&\\
  &\mathsf{Fields}(\mathsf{Cons}^{\mathsf{Vec}}) :=
      \!\cdot,
     (A\! : \! \typeType),(n\! : \! \bN), (m\! : \! \bN), (h\! : \! A),\ (t\! : \! \mathsf{Vec}\,A\,m), (eq\! : \! n = m{+}1)&
\end{flalign*}

In Agda-style notation, a safe head function for vectors can be written:
\begin{flalign*}
  & \mathsf{head} : (A : \typeType) -> (n : \bN) -> \mathsf{Vec}\ A \ (1 + n) -> A&\\
  & \mathsf{head}\ A\ n\ (\mathsf{Cons}\ h\ t) = h &
\end{flalign*}
In \lang this would be defined as:
\begin{flalign*}
&\mathsf{head} : \Pi(A : \typeType) (n : \bN) (x : \mathsf{Vec}\ A\ (n + 1)) \ldotp A &\\
& \mathsf{head} := \lambda A \ldotp \lambda n \ldotp \lambda x \ldotp
  \match{\cdot \envext n \envext x}{\cdot,(n : \bN) , (x : \mathsf{Vec}\,A\,(n+1) )}{A}{
      \\&\qquad \cdot, (m : \bN), (h : A), (t : \mathsf{Vec}\,A\,m)\ldotp \cdot \envext m \envext \mathsf{Cons}\,A\,(m + 1)\, m\,h\,t\,\refl{m+1}
  \branch h} &
\end{flalign*}

That is, the function takes $A$, $n$, and $x$ as type, number, and vector parameters.
It then passes those parameters as the scrutinee of the pattern match,
which is annotated with their types. The annotation is a telescope of types, so it introduces
new names for the scrutinees.
It happens that here we are passing names as the scrutinees, but this need not be the case,
which is why we need new names for them.
The result type is annotated as $A$.
The match has a single branch, with a telescope of
pattern variables $m$, $h$ and $t$ with their types.
The pattern for the branch has $m+1$ as the value for scrutinee $n$ and
$\mathsf{Cons}$ applied to its arguments for scrutinee $x$.
Finally, to the right of $\branch$ is the result for this case,
which is $h$.

 Implicit in this example is the need for the following to hold:
\begin{flalign*}
  & \set{\cdot, (m : \bN) (h : A), (t : \mathsf{Vec}\ A\ m)\ldotp \cdot \envext m \envext \mathsf{Cons}\ A\ (m+1)\ m\ h\ t\ \refl{m+1} }
    \\ & \qquad \covers   \cdot,(n : \bN) , (x : \mathsf{Vec}\ A\ (n + 1) )
\end{flalign*}
That is, the pattern for the single branch needs to cover the scrutinee type.
Deducing that these patterns are a  valid coverage involves seeing that $\refl{m+1}$ constrains the value
of $n$ and that $\mathsf{Nil}$ has an absurd type.

The goal of this paper is to define direct semantics that justify such deductions,
and give a broad framework to define valid coverages.


\section{Categorical Models of \lang}
\label{sec:cwf}
In this section, we translate the syntactic constructs of \lang into
the language of \concept{Categories with Families} (CwFs).
CwFs correspond almost exactly to the syntactic structure of dependent type theory,
but with syntactic substitution replaced by a semantic operation, and with
implicit liftings between contexts made explicit.

\subsection{Background: Categories with Families}

We recapitulate the definition and notation of CwFs, a
categorical model for dependent type theory that follows the syntax
fairly closely.  See
\citet{hofmannSyntaxSemanticsDependent1997} for more details.

Recall that a \concept{family} $X$ is a pair $\pair-{I_X}{\carrier X }$
consisting of a set $I_X$ and an $I$-indexed sequence of sets
$\family-[i \in I]{\carrier X_i}$. A \concept{map} of families $f : X \to
Y$ is a pair $\pair-{f_I}{\carrier f}$ consisting of a function $f_I :
I_X \to I_Y$ and an $I$-indexed sequence of functions $\family-[i \in
  I]{f_i : \carrier X_i \to \carrier Y_{f i}}$. The category $\CatFam$
has families as objects and maps of families as morphisms, with
componentwise identity and composition structure.

A \concept{basic CwF}\/ $\McC$ is a pair $\pair-{\ordinary\McC}{F}$
consisting of a category $\ordinary\McC$ and a functor $F :
\ordinary\McC^{\op} -> \CatFam$.  The functor $F$ packs four pieces of
structure which we'll unpack using the following notation:
\begin{itemize}
\item Objects $\Gamma \in \ordinary\McC$ are \concept{contexts},
  and morphisms $\theta : \Gamma \to \Delta$ are \concept{substitutions}.
\item For every context $\Gamma \in \ordinary\McC$, we denote the
  family $F \Gamma$ by $\pair-{\Ty\Gamma}{{\Tm[\Gamma]{ \_}}}$.  We call
  the elements of its indexing set $\Ty\Gamma$ the \concept{types} in
  context $\Gamma$. For each type in $T \in \Ty\Gamma$, we call the element of
  the component $\Tm[\Gamma]{ T}$ the \concept{terms} of type $T$ in
        context $\Gamma$.
        We omit  $\G$ when it is clear from context.
\item For every substitution $\theta : \Gamma \to \Delta$, we have a
  map of families $F\theta : F\Gamma \from F\Delta$, and we use the
  same notation for both its components $F\theta=
  \pair{\applySub\_\theta}{\applySub\_\theta}$ and call both
  \concept{substitution functions}. The first component is the
  substitution function on \concept{types} $\applySub\_\theta : \Ty\Gamma
  \from \Ty\Delta$. The second component is a sequence of substitution
  functions on the terms of each type $\applySub\_\theta : \Tm[\Gamma]{
    \applySub{T}\theta} \from \Tm[\Delta]{ T}$.
\item The functoriality of $F$ amounts to the following four
  properties, which we call the \concept{substitution lemma} for this CwF,
  where $T$ ranges over $\Ty\Gamma$, $t$ ranges over $\Tm[\Gamma]{ T}$:
  \[
  \applySub{T}{\id[\Gamma]} = T
  \quad
  \applySub{t}{\id[\Gamma]} = t
  \quad
  \applySub{T}{\theta \compose \sigma} = \applySub{(\applySub{T}\theta)}\sigma
  \quad
  \applySub{t}{\theta \compose \sigma} = \applySub{(\applySub{t}\theta)}\sigma
  \tag*{(\textit{where }$\Xi \xto{\theta} \Delta \xto\sigma \Gamma$)}
  \]
\end{itemize}
A basic CwF includes only the bare bones of semantic models for a
dependent type theory. In order to model dependent type theories of
interests we need to equip them with additional structure.

We start with context extension.  Let $\McC$
be a basic CwF, and assume $\ordinary\McC$ has a terminal object $\emptyCtx$.  A
\concept{comprehension structure} $\triple-{\cext}{\p}{\v}$ over
$\McC$ consists of, for each context $\Gamma \in \ordinary\McC$ and
type $T \in \Ty\Gamma$:
\begin{itemize}
\item A context $\Gamma \cext T$, the context $\Gamma$
  \concept{extended by} $T$.
\item A substitution $\p_T : \Gamma \cext T \to \Gamma$, the
  \concept{weakening} of $\Gamma$ by $T$. We say that we
  \concept{weaken} a type or a term by $T$ when we apply the
  corresponding substitution function for the weakening $\p_T$.
\item A term $\v_T \in \Tm[\Gamma \cext T]{T}$, the \concept{variable}
  we extend the context $\Gamma$ with.
\item Moreover, for every substitution $\theta : \Delta -> \Gamma$ and term
  $t \in \Tm[\D]{\applySub{T}{\theta}}$
  there is a unique substitution
  $\ext[: T]{\theta}{t} : \Delta -> \Gamma \cext T$ satisfying:
  \[
  \p_{T} \compose \ext{\theta}{t} = \theta
  \qquad \applySub{\v_{T}}{\ext{\theta}{t}} = t
  \]
  We call this substitution the \concept{extension} of the substitution
  $\theta$ by $t$.
  In the sequel we omit the type ascription and write $\ext\theta
        t$ for $\ext[: T]\theta{t}$ when $T$ is clear from context.
  Likewise, we omit the subscript on $\p$ and $\v$ when clear.
\end{itemize}
A CwF is a basic CwF together with comprehension structure.
For a CwF $\McC = \pair-{\ordinary\McC}{F}$, we refer to ${\ordinary\McC}$
as $\McC$ when doing so will cause no ambiguity.

\subsection{Sections, Slices, and Dependent Types}
\label{subsec:sec-slice}
Let $\McC$ be a category and $\Gamma \in \McC$ an object in it. Recall
the slice category $\McC / \Gamma$ whose objects $\pair{T}{d}$ consist
of an object $T \in \McC$ and a morphism $d : T \to \Gamma$.
Morphisms $f : \pair{T}{d} \to \pair{S}{e}$ in the slice are morphisms $f :
T \to S$ that lift $d$ through $e$, i.e.: $e\circ f = d$. For example,
a section of $\Gamma$ is a morphism out of $\pair{\Gamma}{\id}$ in the
slice $\McC / \Gamma$, i.e., an object $T \in \McC$, and a pair of
morphisms $f : \Gamma \to T$ and $d : T \to \Gamma$ such that $d \circ
f = id$.

We can use the slices of a category $\McC$ to model dependent type
theory by requiring $\McC$ to be locally Cartesian closed (LCCC). The
intuition behind this structure is that the object $\pair Td$ in the
slice $\McC / \Gamma$ represent types $T$ in context $\Gamma$, and $d$
represents the dependency of terms of this type on their context. We
will not recapitulate the LCCC conditions explicitly. We will,
however, spell the induced LCCC structure needed for a CwF, for two
reasons. First, we indicate what we have formalized in
\lean. Second, we rely on this relationship between type families and
slice objects in \cref{sec:sheaf}. It lets us use core results of
sheaf theory to model dependencies in pattern matching.


\begin{lemma}
  \label{lem:slice-equiv}
  $\leanCheck$ Let $\McC$ be a CwF.
  \begin{itemize}
  \item The weakening $\p_{T} : (\Gamma \cext T) \to \Gamma$
    encodes the type $T$ as the slice object $\pair{(\Gamma \cext
      T)}{\p_T}$, in the sense that sections of $\Gamma$ correspond to
    terms in context $\Gamma$:
    \[
    \Tm[\Gamma]{T} \isomorphic
    \Hom[(\McC / \Gamma)]{\vphantom\sum\pair\Gamma\id}{\pair{\Gamma \cext T}{\p_{T}}}
    \]
  \item
  The morphisms in $\McC$
  encode indexed types---for all $\Gamma, \Delta \in \McC$, $T
  \in \Ty{\Gamma}$, and $\theta : \Delta -> \Gamma$:
  \[
  \Tm[\Delta]{ \applySub{T}{\theta}}
  \isomorphic
  \Hom[(\McC/\Gamma)]{\vphantom\sum\pair\Delta\theta}{\pair{\Gamma \cext T}{\p_{T}}}
          \]
  \end{itemize}
\end{lemma}
For $t \in \Tm[\G]{T}$ we write $\overline{t} : \G -> \G \cext T$ for $\ext{id}{t}$, the corresponding $\p$-section.

\subsection{Semantic Type Formers and Closedness}
\label{subsec:tyform}

CwFs give the core structure of type dependency, but give no indication of what
types a model supports. Here we give semantic closedness conditions
which postulate the existence of type and term constructors
corresponding to common features of a dependent type theory.

\subsubsection{Dependent Functions}

We say that a CwF $\McC$ supports dependent functions if it has:

\begin{itemize}
  \item For each $S \in \Ty{\Gamma}$ and $T \in \Ty{\Gamma \cext S}$, a type $\Pi(S,T) \in \Ty{\Gamma}$;
  \item For each $T \in \Ty{\Gamma \cext S} $ and $t \in \Ty{T}$, a term $\lambda(t) \in \Tm{\Pi(S,T)}$;
  \item For each $T \in \Ty{\Gamma \cext S} $, $s \in \Tm{S}$ and $t \in \Tm{\Pi(S,T)}$,
        a term $\mathsf{App}(t,s) \in \Tm{\applySub{T}{\ext{id}{s}}}$;
\end{itemize}
such that the usual structural substitution rules hold, as well as the $\beta$-reduction equality
$  \mathsf{App}(\lambda(t),s) = \applySub{t}{\ext{id}{s}}$.
We could impose a version of an $\eta$ rule, but this is not required for our results.

\subsubsection{Equality}
\label{subsec:cwf-eq}
A CwF $\McC$ supports propositional equality types~\citep{hofmannSyntaxSemanticsDependent1997} if, for every type $T \in \Ty{\Gamma}$, there exists substitution-stable terms as follows:
\begin{itemize}
  \item A type $\mathsf{Id}(T) \in \Ty{\Gamma \cext T \cext \applySub{T}{\p}}$;
  \item A morphism $\mathsf{Refl}_{T} : \Gamma \cext T -> \Gamma \cext T \cext \applySub{T}{p} \cext \mathsf{Id}(T)$,
        such that $\p \circ \mathsf{Refl} = \ext{id}{\v_{T}}$;
  \item For each $S \in \Ty{\Gamma \cext T \cext \applySub{T}{\p} \cext \mathsf{Id}(T)}$,
        a function (on sets)
        ${\mathsf{J}_{T,S} } $ which is in ${ \Tm{\applySub{T}{\mathsf{Refl}_{T}}} -> \Tm{S}}$,
        where, for any $t \in \Tm{\applySub{S}{\mathsf{Refl}_{T}}}$,
        $\applySub{\mathsf{J}(t)}{\mathsf{Refl}_{T}} = t$.
\end{itemize}

We have analogues of \lang terms using substitution. For
$T \in \Ty{\G}$ and $s,t \in \Tm{T}$:
\begin{itemize}
  \item $\mathsf{Id}(T,s,t) := \mathsf{Id}(T)\sub{ \ext{\ext{id}{s}}{t} }\in \Tm{\Gamma}$;
        \item $\mathsf{Refl}_{T}(t) := \applySub{\v}{\mathsf{Refl}_{T}\circ \overline{t}}
  \in \Tm{\mathsf{Id}(T)\sub{\p \circ \mathsf{Refl}_{T} \circ \overline{t}}} = \Tm{\mathsf{Id}(T,t,t)}$.
\end{itemize}

We say that a CwF supports \concept{extensional equality} when
$\Tm{\mathsf{Id}(T,s,t)} \neq \emptyset $ iff $s = t$,
such as in presheaf or set-theoretic models.
Note that this condition does not require  \lang to soundly model an equality reflection rule.
Intensional equality in \lang can be modelled extensionally, so long as
one does not augment \lang with axioms like univalence, which are inconsistent
with equality reflection.
We focus on extensional models, discussing
alternatives in \cref{subsec:disc-intensional}.

\subsubsection{Labelled Variants}
\label{subsec:cwf-labelled-coprod}
Next we define what it means for a category to support \concept{labelled variants},
so that we can interpret datatypes as coproducts of their constructor types.
Labelled variants are simply coproducts with extra structure to mediate the type-context
relationship.
Assume:

\begin{itemize}
  \item a fixed set of type constructors $\textsc{TyCon}$;
        \item for each $I \in \textsc{TyCon}$, a set of data constructors $\textsc{DataCon}_{I}$;
  \item for each $I \in \textsc{TyCon}$, an object $\mathsf{Params}_{I} \in \McC$
        specifying types of arguments to the type constructor;
  \item for each  $D^{I} \in \textsc{DataCon}_{I}$
        a type
        $\mathsf{Fields}_{D^{I}} \in \Ty{\mathsf{Params}_{I}} $
        specifying the types of arguments to the data constructor.
\end{itemize}
The last condition will usually rely on having some sort of dependent pair to encode multiple fields.
Then a CwF supports the labelled variants for $I$ if
  there exists some $\mathsf{TyCon}_{I} \in \Ty{\mathsf{Params}_{I}}$,
   such that for each $\theta : \Gamma -> \mathsf{Params}_{I}$, there exists an isomorphism
        \begin{displaymath}
          \iota : \Gamma \cext \applySub{\mathsf{TyCon}_{I}}{\theta}
          \isomorphic
           \coprod_{i} (\Gamma \cext \applySub{\mathsf{Fields}_{D^{I}_{i}}}{\theta})
          \quad\text{ s.t. }
          \forall i, \p_{\applySub{\mathsf{Fields}_{D^{I}_{i}}}{\theta}} \!=\! \p_{\applySub{\mathsf{TyCon}_{I}}{\theta}} \o \iota^{-1} \o \mathsf{inj}_{i}
        \end{displaymath}

That is, projecting out the context of from the fields' type is the same as
converting into $\mathsf{TyCon}_{I}$ with $\iota$ and then projecting the context.
Then, for each $t \in \Tm{\applySub{\mathsf{Fields}_{D^{I}_{i}}}{\theta}}$,
the arrow $ \iota^{-1} \o  \mathsf{inj}_{i} \o \overline{t}$
is a section of $\p_{\applySub{\mathsf{TyCon}_{I}}{\theta}}$, and hence denotes
a term in $\Tm{\applySub{\mathsf{TyCon}_{I}}{\theta}}$.
We call this term $\mathsf{DataCon}_{D^{I}_{i}}(t)$, since it denotes
the $i$th data constructor applied to field values $t$.

  \subsection{Pattern Matching}
  \label{subsec:cwf-pat}
Here we give a semantic presentation of \lang-style pattern matching. In the CwF
framework, we can follow the definition of \lang.
This serves as a statement of what we need to define pattern matching semantically.
We show how to fulfill those requirements in \cref{sec:sheaf}.

Consider a semantic coverage relation $\covers$
whose elements are sets containing arrows into $\Delta$.
We say $\McC$ supports matching over the semantic coverage $\covers$ if, \underline{for every},
 $\Delta \in \McC$,
   $T \in \Ty{\Delta}$,
   index set $\mathcal{I}$,
   covering $\setj{i}{\theta_{i} : \D_{i} -> \D} \covers \Delta$ for $i \in \mathcal{I}$,
   branch results $t_{i} \in \Tm{\applySub{T}{\theta_{i}}} $ for $i \in \mathcal{I}$,
   and scrutinee $\theta : \Gamma -> \Delta$,
\underline{there exists}
a term $\mathsf{match}_{\seqi{i}{(\Delta_{i},\theta_{i},t_{i})}}(\theta) \in \Tm{\applySub{T}{\theta}} $ such that:
\begin{itemize}
  \item for any $\sigma_{i}  : \Gamma -> \Delta_{i} $,
        if $\theta = \theta_{i} \circ \sigma_{i}$, then $\mathsf{match}_{\seqi{i}{(\Delta_{i},\theta_{i},t_{i})}}(\theta) = \applySub{t_{i}}{\sigma_{i}}$;

  \item The above choice commutes with substitution, i.e., for $\theta_{1} : \Gamma_{1} -> \Gamma_{2}$ and $\theta_{2} : \Gamma_{2} -> \Delta$, we have
    $\mathsf{match}_{\seqi{i}{(\Delta_{i},\theta_{i},t_{i})}}(\theta_{2} \circ \theta_{1}) = \applySub{(\mathsf{match}_{\seqi{i}{(\Delta_{i},\theta_{i},t_{i})}}(\theta_{2}))}{\theta_{1}}$;
\end{itemize}

To see this concretely, consider the $\mathsf{head}$ function from \cref{subsec:vec-example}
in the CwF structure of \lang.
The scrutinee type $\Delta$ is $\cdot , (A : \typeType), (n : \bN), (x : \mathsf{Vec}\ A\ n)$.
The cover is the singleton $\singleton{ \cdot \envext (m + 1) \envext \mathsf{Cons}\ m\ h\ t\ \refl{m+1}  }$,
where syntactic extension $ \envext $ corresponds to $\ext{\_}{\_}$.
This pattern corresponds to $\theta_{1}$.
The pattern context $\D_{1}:= \cdot, (m : \bN), (h : A), (t : \mathsf{Vec}\ A\ m)$,
so the pattern in the cover corresponds to an arrow
$\cdot, (m : \bN), (h : A), (t : \mathsf{Vec}\ A\ m) -> \cdot,(A : \typeType), (n : \bN), (x : \mathsf{Vec}\ A\ n)$.
There is one branch, whose result $t_{1}$ is $h : A$ (which matches the overall result because the result type is not dependent).
Finally, the scrutinees are the variables $\cdot \envext n \envext x\ $, which have context-type $\D$ in the empty context.

The pattern matching condition says that, if $\covers$ is supported by a model of $\lang$,
and $\singleton{ \cdot \envext (m + 1) \envext \mathsf{Cons}\ m\ h\ t\ \refl{m+1}  } \covers \D$,
then there exists a term $\mathsf{match}_{\D_{1}, \theta_{1}, t_{1}}$
such that, for any $m,h,t$, applying the substitution yields $t$, i.e. $[\cdot \envext (m + 1) \envext \mathsf{Cons}\ m\ h\ t\ \refl{m+1}/\D]\mathsf{match}_{\D_{1}, \theta_{1}, t_{1}} = t$.
In the case of our term model, this is given by the pattern match:
\begin{flalign*}
 & \match{x}{\cdot \envext n \envext x\  :\  \cdot,(n' : \bN) , (x' : \mathsf{Vec}\ A\ n' )}{A}{
      \\ &\qquad \cdot, (m : \bN) (h : A), (t : \mathsf{Vec}\ A\ m)\ldotp \cdot \envext (m + 1) \envext \mathsf{Cons}\ m\ h\ t\ \refl{m+1}
  \branch h}
\end{flalign*}
However, in other models, there may not be an obvious way to form
$\mathsf{match}_{\D_{1}, \theta_{1}, t_{1}}$. We provide a general way of forming such a term in
\cref{sec:sheaf}.


\subsection{Model compatibility and soundness}

%
%
%




The correspondence between the type formers we have introduced
and the constructs of \lang is fairly direct, but we must account for some
technical details to make the connection formal.

Not every possible syntactic cover of \lang leads to a non-trivial model.
For example, if $\set{(x : \bZero) \ldotp \mathsf{inl}\ x  } \covers (\bZero + \bN )$,
then we can write an inhabitant for the empty type:
\begin{displaymath}
  \mathsf{bad} : \bZero  := \match{\mathsf{inr}\ 3}{(\bZero + \bN)}{\bZero}{(x : \bZero)\ldotp (\mathsf{inl}\ x) \branch x}
\end{displaymath}

If we have a CwF structure on a category $\McC$ supporting functions, labelled variants, and pattern matching in the sense of \cref{subsec:cwf-pat} with a coverage relation $\covers$,
then we can soundly model \lang in $\McC$ so long as the syntactic $\covers$ is compatible with the semantic $\covers$.

In \cref{fig:pat-model} we define partial translations $\scott{\G} \in \McC$, $\scott{\G |- T} \in \Ty{\G}$,
$\scott{\G |- t : T} \in \Tm[\G]{T}$.
We omit the cases other than pattern matching, as they are standard.
The translation for pattern matches checks if the patterns translate to a semantic cover,
which is well founded because each pattern is syntactically smaller
than the entire match.

\begin{figure}
  \begin{flalign*}
    & \scott{\cdot} = \bOne \qquad \scott{\G,(x : T)} = \scott{\G}\cext \scott{T} \\
    & \scott{\G |- \cdot => \cdot} = !_{\scott{\G}} : \G -> \bOne \qquad \scott{\G |- \env{s} \envext t => \D, (x : T)} =
      \ext{\scott{\env{s}}}{\scott{t}} : \scott{\G} -> \scott{\D}\cext\scott{T} \\
    & \scott{\Gamma |- \match{\env{t_{scrut}}}{\Xi}{T_{motive}}{\seqi{i}{ \Delta_i\ldotp \env{s}_{i} \branch t_{i}}}
      }
     \\ &  \qquad \qquad = \mathsf{match}_{\seqi{i}{(\scott{\Delta_{i}},\scott{\env{s}_{i}},\scott{t_{i}})}}(\scott{\env{t_{scrut}}})
      \in \Tm{\applySub{T}{\scott{\env{t_{scrut}}}}} \\
    &\qquad \textit{ when } \scott{t_{i}} \text{are defined for all $i$ and  }
      \setj{i}{\scott{\env{s}_{i}}}  \covers \scott{\Xi}\\
    & \scott{\Gamma |- \match{\env{t_{scrut}}}{\Xi}{T_{motive}}{\seqi{i}{ \Delta_i\ldotp \env{s}_{i} \branch t_{i}}}
      }
      \textit{ undefined otherwise }
  \end{flalign*}

  \caption{Model of Pattern Matching in \lang}
  \label{fig:pat-model}
\end{figure}
We have the following soundness result, characterizing how the syntactic coverage must correspond to a semantic coverage supporting pattern matching.
\begin{theorem}[soundness]
  \label{thm:soundness}
  If every \lang syntactic cover $\setj{i}{\D_{i} \ldotp \env{s}_{i}} \covers \Xi$ has
   $\scott{\D_{i}}$ and $\scott{\env{s}_{i}}$ defined, and
  $\setj{i}{\scott{\env{s}_{i}}}  \covers \scott{\Xi}$,
  then $\scott{\_}$ is a total mapping on terms/types/environments/contexts that are well typed with respect to $\covers$.
  Moreover, the model is sound with respect to definitional equality.
\end{theorem}
\begin{proof}
  If all covers are compatible, then straightforward induction shows that the undefined case never arises on well-typed terms.
  The argument follows the standard CwF model of type theory,
  except for pattern matching, where the equations from \cref{subsec:cwf-pat} directly
  satisfy \textsc{EqMatch}.
\end{proof}

\section{Coverages and Sheaves to Model \lang}
\label{sec:sheaf}
Theorem~\ref{thm:soundness} lists conditions that ensure we can
soundly interpret \lang.  What categories and
coverages can we find that fulfill our criteria?

%
In this section we connect some core concepts of sheaf
theory to pattern matching.
Specifically, we provide a sufficient condition for when a coverage on a category $\McC$ admits
semantic pattern matching as in \cref{subsec:cwf-pat}.

\subsection{Coverages and Sheaves}
\label{subsec:cover-sheaf}
Alongside our contribution we provide a brief introduction to sheaves and sites for completeness.
Systematic overviews of sheaves are given, for example, by
\citet[C2]{johnstoneSketchesElephantTopos2003}
or \citet{maclaneSheavesGeometryLogic1992}.

\subsubsection{Coverages and Sites}

\begin{wrapfigure}{r}{0.17\textwidth}
  \vspace{-1\baselineskip}
  \centering
  \begin{tikzcd}
    \Xi_j & \G_{i} \\
    \Xi & \D \\
    \arrow["k_j", from=1-1,to=1-2]
    \arrow["h_j", from=1-1,to=2-1]
    \arrow["f_i", from=1-2,to=2-2]
    \arrow["g", from=2-1,to=2-2]
  \end{tikzcd}
    \vspace{-2\baselineskip}
\end{wrapfigure}
A \concept{sheaf-theoretic coverage} on a category $\McC$ is, for each $\Delta \in \McC$,
a set of subsets of $\Hom[\McC]{\_}{\Delta}$, called \concept{covers},
which fulfill the following closure condition~\citep[C2.1.1]{johnstoneSketchesElephantTopos2003}:
\underline{For each} cover $\setj{i \in 1 \ldots n}{f_{i} : \G_{i} -> \D}$,
   and other morphism $g : \Xi -> \D $
\underline{there exists}
 a $\Xi$-cover $\setj{j \in 1 \ldots m}{h_{j} : \Xi_{j} -> \Xi }$
such that, \underline{for each $j$, there exists}
an $f_{i}$ such $g \circ h_{j}$ lifts along $f_{i}$.
That is, there exists a $k_{j}$ making the diagram to the right commute.

This condition is weaker than requiring covers to be closed under
pullback by any arrow. In particular, we do not require $\McC$ to have
all pullbacks. If $\McC$ does have all pullbacks,
one can saturate the coverage so that the property of being a cover
is preserved under pullbacks.

A sheaf-theoretic coverage on $\McC$ gives a coverage $J$ for each
object $\D \in \McC$.  Arrows in a $J$-cover
share a common codomain, so it is clear to which object a cover
belongs.  A \concept{site} $(\McC,J)$ is a category $\McC$ equipped
with a coverage $J$.

We refer to ``sheaf theoretic coverages'' specifically to distinguish them from
coverages in the sense of \cref{sec:syntax,sec:cwf}, \ie the sets of patterns
that we allow. Both denote the sets of morphisms/patterns that cover a given context,
but we don't require pattern coverages to fulfill the sheaf-theoretic closure conditions.
In \cref{subsec:disc-intensional} we show that a constructive syntactic model
cannot fulfill them.



\subsubsection{Sheaves}

The next conceptual tools we need are those of a presheaf and a sheaf.
A \concept{presheaf} $P$ is a functor $P : \McC^{\op} -> \CatSet$.
Sheaf theorists think of presheaves as abstract collection of functions/terms,
with $P\Delta \in \CatSet$ being the set
of functions out of $\Delta$/terms in context $\Delta$.  A \concept{sheaf} is a collection
that `thinks' all pattern matches over every cover uniquely
defines a term. Formulating sheaves precisely involves
multiple nested quantifiers, and we break it down in stages.

For any presheaf $P$ and cover $\seq{\theta_{i} : \Delta_{i} -> \Delta} \in J$, a \concept{matching family}
is a collection $x_{i} \in P(\Delta_{i})$ such that for every $\sigma : \Xi -> \Delta_{i}$
and $\sigma' : \Xi -> \Delta_{j}$, if $\theta_{i} \circ \sigma = \theta_{j} \circ \sigma'$,
then $P(\sigma)(x_{i}) = P(\sigma')(x_{j})$.
I.e., a matching family assigns a $P$ value for all arrows in a cover,
while agreeing in overlapping cases.

An \concept{amalgamation} of a matching family $\seqi{i}{x_{i}}$ over
$\seq{\theta_{i} : \Delta_{i} -> \Delta} \in J$ is an $x \in
P\Delta$ such that $P(\theta_{i})(x) = x_{i}$, i.e., it is a value
in the covered object that is compatible with the matching family.

A \concept{sheaf} on $(\McC,J)$ for a cover $\seq{\theta_{i} :
  \Delta_{i} -> \Delta} \in J$ is a presheaf $P$ such that every
matching family has a unique amalgamation. A presheaf is a $J$-sheaf, or
just a sheaf, when it is a sheaf for each cover in $J$. It is in this
way that a $J$-sheaf is a presheaf that `thinks' all covers admit
pattern-matching.

Let $\yo : {\McC -> (\McC^{\op} -> \CatSet)}$ denote the \concept{Yoneda embedding}
that maps each $\G \in \McC$ to the presheaf $\Hom[\McC]{\blank}{\G}$.
A coverage is \concept{subcanonical} when for every $\Delta \in \McC$, $\yo\Delta$ is a sheaf.
There is a largest such coverage $J_{canonical}$---the \concept{canonical coverage}.
We say a cover is \concept{canonical} when it is in the canonical coverage.
Every representable is a sheaf for a canonical cover, though in \cref{subsec:disc-intensional}
we disprove the converse: there are models
which support pattern matching, so every representable is a sheaf for each allowed pattern,
but where the allowed patterns do not fulfill the necessary conditions
to be a sheaf-theoretic coverage.

\subsection{Pattern Matching via Sheaves}
\label{subsec:sheaf-pat}

The similarity between amalgamation and pattern matching is apparent,
and was informally established by \citet{coquandPatternMatchingDependent1992}:
since morphisms in $\McC$ correspond to substitutions (sequences of terms) in \lang,
the sheaf condition
gives a way to merge arrows (branches) with the same codomain (return type).
However, to model dependent pattern matching, we need to handle the dependency
of the branch result type on the scrutinee's value.
Thankfully, slices give us the tools to model type dependency,
and sheaf theory lets us convert subcanonical coverages on a category to coverages on a slice.
The key properties, which we have mechanized in Lean~\citep{joey_eremondi_2025_14768609},
are as follows (see, e.g. \citet[C2.2.17]{johnstoneSketchesElephantTopos2003}):

\begin{theorem}
  \label{thm:sheaf-slice}
  $\leanCheck$
  If $(\McC,J)$ is a subcanonical site, then for $\Gamma \in \McC$,
  the site $(\McC/\Gamma, J_{\Gamma})$ is subcanonical,
  where we define $\setj{i}{f_{i} : (\Delta_{i}, \theta_{i}) -> (\Xi, \sigma)} \in J_{\Gamma}$
  if and only if $\setj{i}{f_{i} : \Delta_{i} -> \Xi } \in J$.
  In particular,
  if $\setj{i}{f_{i} : \Delta_{i} -> \Gamma}$ is canonical, then $\setj{i}{f_{i} : (\Delta_{i}, f_{i}) -> (\Gamma, id) }$ is too.
\end{theorem}
These properties are related to the \concept{fundamental theorem of topos theory}, which says
that a slice of a sheaf category is equivalent to a category of sheaves over the slice.

We now have what we need to state and prove the main result of this section:
a criterion ensuring that a coverage can model pattern matching.
The following theorem has been mechanized in the Lean 4 theorem prover~\citep{joey_eremondi_2025_14768609}; work is underway to
mechanize the model's soundness and the coverage building rules of \cref{sec:coverages}.

\begin{theorem}
  $\leanCheck$ Consider a CwF $\McC$ and, for each $\D \in \McC$, a relation $\setj{i}{\theta_{i}} \covers \Delta$
  where the $ \theta_{i}$ are disjoint monomorphisms into $\Delta$.
  If all covers in $\covers$ are canonical,
  then $\McC$ supports pattern matching (in the sense of \cref{subsec:cwf-pat}).
\end{theorem}

\begin{proof}
Let $(C, J_{canonical})$ be a canonical site with a CwF structure and a relation $\covers \subseteq J_{canonical}$.
Consider a scrutinee type
 $\Delta : \McC$,
   dependent result type $T \in \Ty{\Delta}$,
   canonical cover $\setj{i}{\theta_{i} : \D_{i} -> \D} \covers \Delta$ of non-overlapping monos,
   branch results $t_{i} \in \Tm{\applySub{T}{\theta_{i}}}$,
   and scrutinee $\theta : \Gamma -> \Delta$.
   To construct
   $\mathsf{match}_{\seqi{i}{(\Delta_{i},\theta_{i},t_{i})}}(\theta) \in \Tm{\applySub{T}{\theta}} $,
   we build $t'_{match} \in \Tm[\D]{\applySub{T}{id}}$ with which we
   compose the scrutinee $\theta$.
   We will show how the sheaf condition corresponds to the pattern match.

  \paragraph{Matches as Arrows}
  Recall from Lemma~\ref{lem:slice-equiv} that
  there is a $\CatSet$-isomorphism
  $ {\Tm[\Delta]{ \applySub{T}{\theta}}
  \isomorphic
  \Hom[(\McC/\Gamma)]{\vphantom\sum\pair\Delta\theta}{\pair{\Gamma \cext T}{\p_{T}}}} $.
  To find a term in $\Tm[\D]{\applySub{T}{id}}$, we use
  an arrow in $\Hom[(\McC/\D)]{(\D, id)}{(\D\cext T, \p)} = \yo(\D\cext T, \p)(\D,id)$.

  \paragraph{Pattern Sets as Slice Covers}
  The patterns $\setj{i}{{\theta_{i} : \D_{i} -> \D}}$ correspond to a canonical $\McC$-cover
  by our premise, so by \cref{thm:sheaf-slice} the cover
   $\setj{i}{\theta'_{i} : (\Delta_{i}, \theta_{i}) -> (\Delta, id) }$ is canonical in $\McC/\D$.

  \paragraph{Branches as Matching Families}
  %
  The branch results of the pattern match form a matching family
  for $\yo((\Gamma\cext T, \p))$.
  Our branches are $\seqi{i}{t_{i} \in \Tm[\D_{i}]{\applySub{T}{\theta_{i}}}}$.
  By \cref{lem:slice-equiv}, this family yields a sequence
  $\seqi{i}{ x_{i} \in \Hom[(C/\D)]{(\Delta_{i}, \theta_{i})}{(\Delta \cext T, \p)} }$,
  i.e., $\seqi{i}{ x_{i} \in \yo((\Delta \cext T, \p))((\Delta_{i}, \theta_{i})) }$,
  which is a matching family for the presheaf $\yo(\Delta \cext T, \p)$
  and the cover $\setj{i}{{\theta_{i} : (\Delta_{i}, \theta_{i}) -> (\D, id)}}$.



  \paragraph{Amalgamating Branches}

  Because the cover is canonical, then $\yo(\Delta \cext T, \p)$ is a sheaf for it.
  The sheaf condition states that the above matching family
  has an amalgamation $x \in \yo(\Delta \cext T, \p)(\D,id)$,
  such that $\theta_{i} \o x = x_{i}$.
  So \cref{lem:slice-equiv} yields a term $t'_{match} \in \Tm{\applySub{T}{id}}$ such that
  $\applySub{t'_{match}}{\theta_{i}} = t_{i}$.

  \paragraph{Equations and the Scrutinee}
  Finally, given a scrutinee $\theta : \G -> \D$, we choose $\applySub{t'_{match}}{\theta}$ as
  $\mathsf{match}_{\seqi{i}{(\Delta_{i},\theta_{i},t_{i})}}(\theta) \in \Tm{\applySub{T}{\theta}} $.
  It is in $\Tm[\G]{\applySub{T}{\theta}}$, so it has the correct type.
  It satisfies the requisite equations. Indeed,
  since $\applySub{t'_{match}}{\theta_{i}} = t_{i}$, whenever $\theta = \theta_{i} \o \sigma_{i}$
  for some $\sigma_{i}$,
  we have
  \begin{math}
    {\applySub{t'_{match}}{\theta}} = \applySub{t'_{match}}{\theta_{i} \o \sigma_{i}}
    = \applySub{(\applySub{t'_{match}}{\theta_{i}})}{\sigma_{i}} = \applySub{t_{i}}{\sigma_{i}}
  \end{math}
  just as \cref{subsec:cwf-pat} requires. For substitution,
  given $\theta = \theta_{2} \o \theta_{1}$, we have
  \begin{math}
    \applySub{t'_{match}}{\theta_{2} \o \theta_{1}}
    = \applySub{(\applySub{t'_{match}}{\theta_{2}})}{\theta_{1}} =
    \applySub{\mathsf{match}_{\seqi{i}{(\Delta_{i},\theta_{i},t_{i})}}(\theta_{2})  }{\theta_{1}}
  \end{math}.
\end{proof}

  The above construction lets us model pattern matching for any canonical cover,
  where the motive type corresponds to a representable sheaf.
  If $J$ is subcanonical, then every representable is a sheaf, so we can
  define dependent pattern matching for any motive type.
  Moreover, the canonical coverage contains the covers from every
  subcanonical coverage, so it suffices that each allowed pattern
  set is a canonical cover.
  The disjointness and injectivity conditions ensure that the branches of a match follow
  the sheaf-theoretic definition of a matching family.
  One could instead require that branches agree on their overlap~\citep{cockxOverlappingOrderIndependentPatterns2014},
  which is suited to matching on real numbers~\citep{shermanComputableDecisionMaking2018}.

  We conclude this section by recalling a property
  characterizing subcanonical covers~\citep[C2.1.11]{johnstoneSketchesElephantTopos2003},
  which we will utilize in \cref{sec:coverages}.
\begin{theorem}
\label{thm:effective-epi}
  A set of arrows $\setj{i}{\theta_{i} : U_{i} -> U}$ is canonical for $\McC$
  if and only if, for every $\sigma : V -> U$ and every object $T \in \McC$,
  the presheaf $\yo(T)$ is a sheaf for the pulled back family $\setj{i}{\sigma^{*}  \theta_{i} : \sigma^{*} U_{i} -> V }$.
\end{theorem}
We can use this theorem to form basic subcanonical coverages.
To build new coverages from old ones, we employ
\concept{saturation conditions:} operations on coverages which do not change which presheaves are sheaves for that coverage.
This is of interest to us because
the canonical coverage is invariant under every saturation:
it is already the largest possible subcanonical coverage, so no covers can be added
without changing its notion of sheaf.
The next section gives several examples of useful saturation conditions.

\section{Tools for Building Coverages}
\label{sec:coverages}

In this section, we take the abstract canonicity condition
and derive concrete rules for forming canonical covers.
This section justifies the idea that \lang
 can begin with a basic set of covers for each type
and obtain a language of patterns by allowing nesting,
variables, and multiple scrutinees.
We provide base coverages, along with composition rules for building complex coverages from simpler ones.
This recreates commonly
supported features of dependent pattern matching: variables,
constructors for labelled variants, pruning absurd branches, and matching on $\mathtt{refl}$ with inaccessible (dot) patterns.
In \cref{subsec:discussion-patsyn} we discuss potential novel coverages that can
be supported using coverage semantics.

\subsection{Identity and Isomorphism}
\label{subsec:id-iso}
The most basic canonical covers are singletons consisting of an isomorphism.
If two contexts are isomorphic, then moving from one to the
other covers all cases.

\begin{theoremEnd}[apxproof]{lemma}
  \label{lem:iso-cover}
  A presheaf is a sheaf for a singleton cover containing an isomorphism $\iota : \Gamma \cong \Delta$, so every isomorphism
  is a canonical singleton cover.
 \footnote{
In a category with pullbacks, sheaves may be defined for
{Grothendieck pretopologies}, in which all isomorphisms definitionally yield singleton covers.
}
\end{theoremEnd}
\begin{proofEnd}
  A matching family for $\singleton{\iota}$ is just an element of $x \in P(\Gamma)$.
  We can use $P(\iota^{-1})(x)$ as our amalgamation, and since $P$ is a functor,
  $P(\iota)(P(\iota^{-1})(x)) = P(\iota^{-1}\circ \iota)(x) = P(id)(x) = x$.
  Every presheaf is thus a sheaf for $\singleton{\iota}$, so every representable is, so the largest
  coverage with all representables as sheaves must contain all isomorphisms.
\end{proofEnd}

As a consequence, identity arrows are in singleton canonical covers:
\begin{corollary}
  \label{cor:id-cover}
  The singleton cover $\singleton{id : \Gamma -> \Gamma}$ is canonical for any category, so a pattern consisting entirely of variables
  $x_{1}, x_{2}, \ldots x_{n}$
  can be safely included in any coverage for \lang.
\end{corollary}

Supporting identity arrows is
the bare minimum we need for pattern matching.
They are the base case out of which other patterns are built, where no
discrimination or computation happens at all.
Likewise, a catch-all pattern, commonly written as an underscore `$\blank$',
is just an unnamed variable that does not occur in the right-hand side of the branch.

More generally,
the canonical coverage contains all isomorphisms.
So we can devise a sound semantics for \lang where any isomorphism
is a valid cover of a type.
This allows for operations such as rearranging variables in a dependency-respecting
way or re-bracketing nested sums and products.

Functions written by the programmer can even be used as patterns if they are isomorphisms,
opening the door for user-defined views into a type.
Of course, the existence of a sound semantics does not guarantee a
language we can actually implement. Checking whether a term is a definitional
isomorphism is undecidable without being explicitly given its inverse.
Moreover, depending on how extensional the model is, there may be terms that are
isomorphisms in the model, but are not definitional isomorphisms in \lang.

\subsection{Coproducts and Wadler Views}
With variables as patterns, the next primitive patterns we need are constructors
for a datatype. As we saw in \cref{subsec:cwf-labelled-coprod}, one way to model
datatypes is with labelled variants.
So if $\McC$ has coproducts, we can model datatype constructors as injections
into a coproduct context, and we can amalgamate branches that match
on all the constructors of a datatype using the universal property of a coproduct.

Labelled variants are only coproducts up to isomorphism, but we have seen that isomorphisms
are always singleton covers, and below in \cref{subsec:cover-comp} we see that composition preserves
canonical covers.
So it suffices to consider coproducts directly.
Unfortunately, in an arbitrary category, coproduct injections are not guaranteed to form a canonical cover.
\cref{thm:effective-epi} requires each representable to be a sheaf for the pullback
of every cover, so we need coproducts to be stable under pullback, i.e. the pullback
of a coproduct is the coproduct of pullbacks.
Thankfully, pullback stability of coproducts holds if $\McC$ is $\CatSet$, a presheaf category,
a topos, or any other locally cartesian closed category (LCCC).
We already want $\McC$ to be LCCC in order to support dependent functions.

\begin{theoremEnd}[apxproof]{theorem}
Suppose $\McC$ has all pullbacks and that coproducts are disjoint and stable under pullback.
Then
$\setj{j}{ \inj_{j} : \Delta_{i} -> \coprod_{i \in I}\Delta_{i} }$ canonically cover $\coprod_{i \in I}\Delta_{i}$.
\end{theoremEnd}
\begin{proofEnd}
  Recall that for $\setj{j \in I}{ \inj_{j} : \Delta_{j} -> \coprod_{i \in I}\Delta_{i} }$ to be in the canonical coverage,
  for every $\Gamma$
  and every $g : \Xi -> \coprod_{i \in I}\Delta_{i}$,
  $\yo(\Gamma)$ should be a sheaf for $\setj{j \in I}{ g*\inj_{j} }$.
  So for each $j$ there is pullback diagram that looks like this:

  \begin{center}
    \begin{tikzcd}
      g^{*} ( \Delta_j)  \rar{} \dar{g^{*} (\inj_j)} & \Delta_j \dar{\inj_j} \\
      \Xi \rar{g} & \coprod_{i \in I} \Delta_i
    \end{tikzcd}
  \end{center}

  Since the identity always has a pullback and $\coprod_{i \in I} \inj_i = id$, we have the following pullback diagram:
  \begin{center}
    \begin{tikzcd}
      \Xi \rar{g} \dar{id} & \coprod_{i \in I}\Delta_i \dar{\coprod_{i \in I} \inj_i} \\
      \Xi \rar{g} & \coprod_{i \in I} \Delta_i
    \end{tikzcd}
  \end{center}

  Since $\McC$ preserves coproducts under pullback, this means that there is an isomorphism
  $\iota : \Xi \cong \coprod_{i \in I} (g^{*}\inj_{i})$, and we have the following pullback diagram:

  \begin{center}
    \begin{tikzcd}
      \coprod_{i \in I} (g^{*} \Delta_i) \rar{ g \circ \coprod_{i \in I}(g^{*}\inj_i)} \dar{ \coprod_{i \in I} (g^{*} \inj_i)} & \coprod_{i \in I}\Delta_i \dar{\coprod_{i \in I} \inj_i} \\
      \Xi \arrow[u, bend left, "\iota"] \rar{g} & \coprod_{i \in I} \Delta_i
    \end{tikzcd}
  \end{center}

  Let's return to the sheaf condition.
  Consider some $\setj{j \in I}{x_{j} : g^{*}\Delta_{j} -> \Gamma}$.
  For $\yo(\Gamma)$ to be a sheaf for $\setj{ g^{*} \inj_{j}}$,
  there should be a unique $x : \Xi -> \Gamma$ such that
  $x \circ g^{*}\inj_{j} = x_{j}$.
  The universal property of the coproduct gives us a unique $\coprod_{i \in I}(x_{i}) : \coprod_{i \in I}(g^{*}\Delta_{i}) -> \Gamma$.
  We can then set $x := \coprod_{i \in I}(x_{i}) \circ \iota$.

  First we must prove that $x$ respect each $\inj_{j}$.
  We have $x \circ g^{*}\inj_{j} = \coprod_{i \in I}(x_{i}) \circ \iota \circ g^{*}\inj_{j}$, so
  it suffices to show that $\iota \circ  g^{*}\inj_{j} = \inj_{j} : g^{*}\Delta_{j} -> \coprod_{i \in I} g^{*}\Delta_{i}$.
  We have $id_{\coprod_{i \in I} g^{*}\Delta_{i}} = \iota \circ \coprod_{i \in I}(g^{*}\inj_{i}) = \coprod_{i \in I}(\iota \circ g^{*}\inj_{i})$,
  so $\inj_{j} = id_{\coprod_{i \in I} g^{*}\Delta_{i}} \circ \inj_{j} = \coprod_{i \in I}(\iota \circ g^{*}\inj_{i}) \circ \inj_{j} = \iota \circ g^{*}\inj_{j}$.

  All that remains is to show the uniqueness of $x$.
  Suppose we have $y : \Xi -> \Gamma$ such that $y \circ g^{*}\inj_{j} = x_{j}$.
  Then $y = y \circ \coprod_{i \in I}(g^{*}\inj_{i}) \circ \iota = \coprod_{i \in I} (y \circ g^{*}\inj_{i}) \circ \iota = \coprod_{i \in I}(x_{i}) \circ \iota = x$.
\end{proofEnd}

The immediate result of this theorem is that for every inductive type, the constructors
are covering for that type, so long as the inductive type is modelled as the labelled variants
of its constructor types.
However, it is important to realize that this theorem applies for any decomposition
of a type into the coproduct of other types, regardless of whether the injections
correspond to constructors or not.
Such a coverage introduces the possibility for views as introduced by \citet{wadlerViewsWayPattern1987}, which act as first class pattern synonyms:

\begin{corollary}
  \label{cor:coprod-cover}
  Consider a finite $I : \typeType{}$, a family $S : I -> \typeType{}$ and
  an indexed function $f : (i : I) -> S\ i -> T$ in \lang.
  Suppose we have a category $\McC$  with a CwF model of \lang supporting pattern matching as in \cref{subsec:cwf-pat}
  over a coverage $\_\covers\_$.
  If
  $\scott{T} \cong \scott{\Sigma(t : T)(i : I)(s : S\ i)\ldotp t =_{T} f\ i\ s )}$,
  then there is also a model of \lang with coverage $(\_\covers\_ \cup \setj{i}{ f\ i})$
\end{corollary}

That is, if $T$ is isomorphic to the sum of some types $S\ {i}$ over finite $i$,
we can safely match on a value from $T$, where the $i$th pattern
is the $S\ i$ value that it corresponds to,
regardless of whether $T$ is defined as an inductive type or the $S\ i$
are its constructor types.

\subsection{Nesting and Composition}
\label{subsec:cover-comp}

With variables and constructors as primitive covers,
we now need a way to combine them to build more complex covers.
 Adding the composition of different coverages
does not change their sheaves.
Every cover has the same sheaves as the sieve it generates, i.e.,
the closure of the cover under precomposition with any arrow in $\McC$~\citep[C2.1.3]{johnstoneSketchesElephantTopos2003}.
So canonical covers are closed under composition:

\begin{theoremEnd}[apxproof]{theorem}
 \label{thm:cover-comp}
  For a cover $J$ of $\McC$,
  If $\setj{i}{f_{i} : \Delta_{i} -> \Delta} \in J$,
  and for each $i$, we have $\setj{ij}{g_{ij} : \Delta_{ij} -> \Delta_{i}} \in J$,
  then the sheaves of $(\McC,J)$ are identical to the sheaves of
  $(\McC, J \cup \setj{i}{f_{i} \o g_{ij} : \Delta_{ij} -> \Delta })$.
  So if $\setj{i}{f_{i} : \Delta_{i} -> \Delta }$ is canonical,
  and for each  $i$, $\setj{ij}{ g_{ij} : \Delta_{ij} -> \Delta_{i}}  $ is canonical,
  then $ \setj{ij}{f_{i} \o g_{ij} : \Delta_{ij} -> \Delta }$ is canonical.
\end{theoremEnd}
This allows patterns to be nested:
if  $\setj{i}{f_{i}}$ are covering for $\Delta$, and for each set of variables
in those covering patterns, $\setj{ij}{g_{ij}}$ is covering, then we can case split each variable
in the $f_{i}$ into $j$ cases corresponding to the $g_{ij}$ patterns, and the entire resulting
set can still be covering.
This property gives semantic justification for the case-split operation of the Agda and Idris editor
modes, where the programmer selects a variable in a pattern match,
and the pattern containing the variable is replaced by the sequence of patterns
that has each possible constructor application in place of the variable.

When we combine the closure of the canonical coverage under identity (isomorphisms),
sum injections, and composition, we can recreate dependent pattern matching
on non-indexed datatypes. However, we see in the next sections that the language of coverages
also gives us the tools to handle indexing.

\subsection{Pruning Absurd Contexts}
\label{subsec:absurd-cover}

If we allow ourselves some extensionality, then we can use sheaves to
model absurd branches and empty cases.
Suppose that $\McC$ has an initial object $\bZero$,
with a unique arrow $0_{\G} : \bZero -> \G$
for every $\G$.
The initial context denotes an empty or absurd context, since we can derive a term of any type from it.
It turns out that we do not ever need to include branches for patterns
whose contexts are empty.
If we can amalgamate for a cover where one pattern has an empty context,
we can amalgamate for the same cover with that arrow deleted, since any matching
family for the smaller cover can be turned into one for the larger cover
by adding the unique arrow out of the initial context.

\begin{theoremEnd}[apxproof]{theorem}
  \label{thm:absurd-cover}
  Let $c$ be a canonical cover with $\theta : \D -> \G \in c$.
  If there exists an arrow $\iota : \D -> \bZero$, then $c \setminus \singleton{ \theta }$ is canonical.
\end{theoremEnd}
\begin{proofEnd}
  The arrow $\iota$ is necessarily an isomorphism by the uniqueness of initial objects,
  so we can use \cref{thm:cover-comp} replace $\theta $ with $0_{\Gamma}$ and still have a canonical cover.
  We need to show that for any $g : \Xi -> \Gamma$ and any motive $M \in \McC$, $\yo(M)$ is a sheaf for $g^{*}(c \setminus \singleton{\theta})$.
  We have $g^{*}(0_{\Gamma}) = 0_{\Xi}$ by the uniqueness of initial arrows
  So in a matching family for $\yo(M)$, the only possible value in $\yo(M)(g^{*}(0_{\Gamma})) $ is $0_{M}$.
  So given a matching family for $g^{*}(c \setminus \singleton{\theta})$, we can always construct one for $g^{*}(c)$,
  which we know is a canonical cover, and hence we can amalgamate for.
\end{proofEnd}

This property mirrors how Agda and Idris allow for the omission of empty cases.
In some cases, these languages only allow branch right-hand sides to be removed after the programmer specifies an empty pattern,
marking which part of the scrutinee has an impossible type.
We view this empty pattern as a syntactic aid
to tell the type checker when a context is isomorphic to the initial context,
so we do not directly model the empty pattern.

Like our assumption about equality, the condition of having a (strong) initial object does not hold in the term model,
since not all eliminations of the empty type are definitionally equal. So long
as \lang does not contain any axioms that specifically distinguish empty eliminations,
 our model is still sound.

\subsection{Propositional Equality}
\label{subsec:cover-eq-var}

Since isomorphisms are always canonical singleton covers (\cref{lem:iso-cover}), we can create
a coverage for a sufficiently-extensional equality type.
\begin{corollary}
  \label{cor:refl-cover}
  If
  $ { \ext{\v}{\ext{\v}{\mathsf{Refl}_{A}}} : \Gamma \cext A -> \Gamma \cext A\cext \applySub{A}{\p} \cext \mathsf{Id}(\applySub{A}{\p^{2}},\v,\applySub{\v}{\p}) }$ is an isomorphism,
  then $\singleton{\ext{\v}{\ext{\v}{\mathsf{Refl}_{A}}}}$ is canonical.
\end{corollary}
In more readable, non CwF notation: if $A$ is isomorphic to $\Sigma(x : A)(y:A) \ldotp x =_{A} y$
in the model via the projections,
then $\singleton{ (x,x,\refl{x}) }$ can be a singleton cover for
$\Sigma(x:A)(y:A)\ldotp x =_{A} y$.
Such an isomorphism holds if $\McC$ has extensional equality,
since it asserts that
there is a unique, internally constructible proof of equality between two equal terms.

For a dependent match targeting
${(x \!:\! A),(y\!:\!A), (pf \!:\! x \!=_{A}\! y) |- P(x,y,pf) \!:\! \typeType{}}$,
the above cover only requires we provide a branch result with type $P(x,x,\mathsf{Refl}_{A})$.
The variable $y$ was replaced by $x$ in the goal type.
This captures ``inaccessible'' or ``forced''
patterns~\citep{norellDependentlyTypedProgramming2009},
known as dot-patterns in Agda and Idris. By matching on the propositional equality,
we work with refined information about the context.
Here, we match $y$ against the variable $x$ rather than
a constructor. No branching or discrimination
happening, since the cover is a singleton. Rather, $x$ is the only possible value for $y$ given the equality proof.
Agda writes  $.x$ to express this pattern.
\Cref{subsec:eq-pullback} extends this to equality
proof between arbitrary terms, rather than variables.


\subsection{Pullbacks and Unification}
\label{subsec:base-change}

We have seen that the canonical coverage includes isomorphisms and injections and that
it allows for composition. However, the definition of a sheaf-theoretic
coverage enables stability under pullback: for any coverage, a sheaf for that coverage
is still a sheaf if we add the pullback of any cover by any arrow.
Closure under pullback is the key condition that separates a coverage from a set of arrows.
The definition in \cref{subsec:cover-sheaf} is presented in the style of \citet[C2]{johnstoneSketchesElephantTopos2003},
in a general way that does not assume the existence of pullbacks. However, when each
arrow in a cover has a pullback along some morphism, we get the following saturation condition:

\begin{theorem}
  \label{thm:base-change}
  For a cover $J$ of $\McC$,
  if $\setj{i}{\theta_{i} : \Delta_{i} -> \Delta} \in J$,
  and $g : \Gamma -> \Delta$,
  where the pullback of each $\theta_{i}$ along $g$ exists,
  then
  $J \cup \singleton{\setj{i}{g^{*}\theta_{i} : g^{*} \Delta_{i} -> \Gamma}}$
  has the same sheaves as $J$.
  So
  if $\setj{i}{\theta_{i} : \Delta_{i} -> \Delta} $ is canonical,
  then so is
  $ \setj{i}{g^{*}\theta_{i} : g^{*} \Delta_{i} -> \Gamma} $.
\end{theorem}

This abstract property, known as \concept{stability under base change},
can be exploited
to build interesting covers.

\subsubsection{Context Extension}
\label{subsec:cover-ext}
Base change lets us add new scrutinees to a pattern match.
In any CwF, for $\theta : \Gamma -> \Delta$ and $T \in \Ty{\Delta}$, pulling back by $\p : \Delta \cext T -> \Delta$
yields a morphism $\ext{\theta \circ \p}{\v} :  \Gamma \cext \applySub{T}{\theta} -> \Delta \cext T$~\citep{hofmannSyntaxSemanticsDependent1997}.
Combining this with \cref{thm:base-change} gives:
\begin{theorem}
For canonical $\setj{i}{ \theta_{i} : \Delta_{i} -> \Delta } $  and a type $T \in \Ty{\D}$, there is also a canonical cover
${\setj{i}{ \ext{\theta_{i} \circ \p}{\v}:
    \Delta_{i} \cext \applySub{T}{\theta_{i}} -> \Delta\cext T } }$.
\end{theorem}
So we can build a covering pattern for a context $\Delta$ and immediately obtain a covering context
on $\Delta \cext T$ by appending a new variable $\v \in \Tm{\applySub{T}{\theta_{i}}}$  to each pattern in the cover.
In a dependent match the new variable might have a different type in each branch:
$T$ may be indexed by variables in $\Delta$,
but each $\theta_{i}$ is a value for $\D$ with variables from $\D_{i}$.
Further case-splitting on the newly introduced variable can be achieved using composition
\ala \cref{subsec:cover-comp}.

\subsubsection{Matching on Equality}
\label{subsec:eq-pullback}



Suppose that for some $\G \in \McC$ and $T \in \Ty{\Gamma}$, and that:
$  \singleton{\mathsf{Refl}_{T} : \Gamma \cext T -> \Gamma \cext T \cext \applySub{T}{\p} \cext \mathsf{Id}(\applySub{T}{\p^{2}}, \v, \applySub{\v}{\p})}
  $ is canonical.
As in \cref{subsec:cover-eq-var}, such a property holds for an extensional model.
With such a cover on equality, we can apply the base change theorem, the CwF laws, and the properties of equality (\cref{subsec:cwf-eq}) to obtain a cover on contexts containing equalities
by matching:

\begin{theoremEnd}[apxproof]{theorem}
  Suppose $\McC$ has all pullbacks.
  Consider $\G \in \McC$ with $T \in \Ty{\G}$, $t_{1},t_{2} \in \Tm{T}$.
  Then pulling back $\mathsf{Refl}_{T}$ by $\ext{\ext{\overline{t_{1}\sub{\p}}}{t_{2}\sub{\p}}}{\v}$ yield a context $\D$
  and an arrow
  $\ext{\theta}{\mathsf{Refl}_{T}(t_{12})}$, where $\theta : \D -> \G$, $t_{12} \in \Tm{T\sub{\theta}}$,
  and $\applySub{t_{1}}{\theta} = \applySub{t_{2}}{\theta} = t_{12}$.
  Moreover, if $\singleton{\mathsf{Refl}_{T}} $ is canonical, then so is the cover
  $\singleton{\ext{\theta}{\mathsf{Refl}_{T}(t_{12})} : \D -> \G \cext \mathsf{Id}(T,t_{1},t_{2})} $.
\end{theoremEnd}
\begin{proofEnd}
  Let $\D := \ext{\ext{\overline{t_{1}\sub{p}}}{t_{2}\sub{p}}}{\v}^{*} \Gamma \cext T$.
  Both arrows out of $\D$ in the pullback square are into extended contexts, so they
  are necessarily equal to extended morphisms. So our pullback square looks like this
  for some $\theta, \sigma: \D -> \G $, $t_{Id} \in \Tm{\mathsf{Id}(T,t_{1},t_{2})\sub{\theta}}$ and $t_{12} \in \Tm{\applySub{T}{\sigma}}$:
\begin{center}
  \begin{tikzcd}
    \Delta \rar{\ext{\sigma}{t_{12}}} \dar{\ext{\theta}{t_{Id}}} & \Gamma \cext T \dar{\mathsf{Refl}_T}\\
    \Gamma \cext \mathsf{Id}(T,t_1,t_2) \rar{\ext{\ext{\overline{t_1}}{t_2}}{\v}} & \G \cext T \cext \applySub{T}{\p} \cext \mathsf{Id}(T)
  \end{tikzcd}\\
\end{center}
Applying our CwF laws we get
  $\ext{\ext{\ext{\theta}{\applySub{t_{1}}{\theta}}}{\applySub{t_{2}}{\theta}}}{t_{Id}} = \mathsf{Refl}_{T}\circ \ext{\sigma}{t_{12}} $.
  Then precomposing with $\p$ on both sides yields
  $\ext{\ext{\theta}{\applySub{t_{1}}{\theta}}}{\applySub{t_{2}}{\theta}} = \ext{\ext{\sigma}{t_{12}}}{t_{12}}$,
  so we have $\theta = \sigma$ and $\applySub{t_{1}}{\theta} = t_{12} = \applySub{t_{2}}{\theta}$.
  Likewise, applying both sides as substitutions to $\v$ gives $t_{Id} = \applySub{\v}{\mathsf{Refl}_{T}\circ \ext{\sigma}{t_{12}} } = \mathsf{Refl}_{T\sub{\theta}}(t_{12}) $
  by the previous equations and the stability of $\mathsf{Refl}$ under substitution.
 \je{TODO stability of Refl under subst}

  Finally, we have $\singleton{\ext{\theta}{\mathsf{Refl}_{T}(t_{12})}}$ as a canonical cover by the base change theorem.

\end{proofEnd}

\begin{wrapfigure}{r}{0.55\textwidth}
  \begin{tikzcd}[column sep=small, style = {font = \footnotesize}]
    \Delta  \dar{\theta \envext \refl{t_{12}}}  \rar{\theta \envext t_{12}} & \Gamma,(x : T)\dar{(x \envext x \envext \refl{x})}\\
    \Gamma,(pf\!:\! x\!=_T\! y) \rar{(t_1 \envext t_2 \envext s)} & \Gamma, (x\!:\!T),(y\!:\!T),(x\! =_T\! y)
\end{tikzcd}
  \end{wrapfigure}
For the intuition behind this, consider the pullback square to the right, translated to \lang-style notation for clarity.
First, because the square commutes, we know that $\theta$ is a substitution
that equates $(t_{1}, t_{2}, pf)$ and $(x, x, \refl{x})$,
i.e., it is
a \concept{unifier} of $t_{1}$ and $t_{2}$.
Since a pullback is a limit, it is universal, so any other unifiers
for $t_{1}$ and $t_{2}$ necessarily factor through $\theta$.
Thus, it is the \concept{most general unifier} for $t_{1}$ and $t_{2}$.
This is precisely what the usual rule for pattern matching on equality uses:
it unifies the two sides of the equality, treating syntactic variables
as unification variables, and generates a substitution that is then
applied to the goal.
The context $\Delta$ consists of the variables that were in common between $t_{1}$ and $t_{2}$
which remain free in the unification $t_{12}$.
In the case that $t_{1}$ and $t_{2}$ do not unify, then the pullback
is an arrow out of an initial context, and the branch can be omitted completely
(because absurd covers can be omitted, as in \cref{subsec:absurd-cover}).


\section{Example: Folding Without a Starting Value}
\label{sec:example}

We now have specified everything we need (sans recursion) for feature-parity with
the original presentation of dependent pattern matching by \citet{coquandPatternMatchingDependent1992}.
Our sheaf-centric view generalizes the elaboration process of \citet{goguenEliminatingDependentPattern2006},
but directly within the model instead of as a syntactic elaboration.
Constructors for a coproduct form a cover and variables form a singleton cover, acting as the basis from which
other covers are generated. Covers can be composed, extended, refined by matching on an equality,
or pared down by pruning absurd branches.
Indexed data types can be handled using fording and matching on equality.

To see a non-trivial example of how to build a cover in the canonical coverage, consider the $\mathsf{foldr}_{1}$ function found in the Agda standard library~\citep{documentationforagdaDataVecBase2024}.
\begin{flalign*}
  & \mathsf{\mathsf{foldr}}_{1} : (A -> A -> A) -> \mathsf{Vec}\,A\,(\mathsf{suc}\,n) ->  A &\\
  &\mathsf{\mathsf{foldr}}_{1}\,f\,(\mathsf{cons}\,x\,\mathsf{nil})\, = x \ \mid\  \mathsf{\mathsf{foldr}}_{1}\,f\,(\mathsf{cons}\,x\, (\mathsf{cons}\,y\, ys)\, = f\,x\,(\mathsf{foldr}_{1}\,f\,(\mathsf{cons}\,y\,ys))& &
\end{flalign*}
Because the argument vector has length at least one, the case for $\mathsf{nil}$ can be omitted.
The base case is then a vector of length one, and the inductive case is a vector
of length two or more.

Assume an extensional CwF model of \lang in a LCCC $\McC$ where inductive types are labelled variants.
We show how the patterns for $\mathsf{foldr}_{1}$ are in the canonical coverage for $\McC$,
and hence $\mathsf{foldr}_{1}$ can be modelled.
Note that because labelled variants are defined in terms of isomorphism,
we do not preclude initial algebra semantics for modelling the self-reference part of inductive types.

\subsection{Translating to \lang}

First, we translate the function to \lang-style by making the length argument explicit and replacing the indexed constructors with ones taking explicit equality proofs.
The datatype becomes:
\begin{flalign*}
  &\mathtt{data}\ \mathsf{Vec}\ (A : \typeType{}) : (n : \bN) -> \typeType\ \mathtt{where}\\
  & \qquad \mathsf{nil} : (n = 0) -> \mathsf{Vec}\ A\ n\\
  & \qquad
  \mathsf{cons} : (m : \bN) -> A -> \mathsf{Vec}\ A\ m -> n = m + 1 -> \mathsf{Vec}\ A\ n.
\end{flalign*}
We also abstract out the recursive calls, since we have not included them in \lang
and have not required that
our model category $\McC$ support them.
Despite using recursion, $\mathsf{foldr}_{1}$ is an ideal example because it
is not contrived, and uses all the main saturation conditions we developed in \cref{sec:coverages}.
Since $\mathsf{foldr_{1}}$ is decreasing in the length of the lists,
its recursion can be modelled with well-known techniques orthogonal to our contribution.

\begin{flalign*}
  & \mathsf{\mathsf{foldr}}_{1} : (n : \bN) -> (A -> A -> A) -> \mathsf{Vec}\,A\,(\mathsf{suc}\,n) \\
  & \qquad \qquad -> (\mathsf{self} : (A -> A -> A) -> \mathsf{Vec}\,A\,(\mathsf{suc}\,n) -> A ) ->  A &\\
  &\mathsf{\mathsf{foldr}}_{1}\,0\,f\,(\mathsf{cons}\,0\, x\,(\mathsf{nil}\,\mathsf{Refl})\,\mathsf{Refl})\,\mathsf{self} = x &\\
  &\mathsf{\mathsf{foldr}}_{1}\,(m + 1)\,f\,(\mathsf{cons}\,(m + 1)\,x\, (\mathsf{cons}\,m\,y\, ys\,\mathsf{Refl})\,\mathsf{self} \\ & \qquad \qquad = f\,x\,(\mathsf{self}\,f\,(\mathsf{cons}\,y\,ys\,\mathsf{Refl})\,\mathsf{Refl})&
\end{flalign*}

\subsection{Building the Coverage}

Using \lang notation rather than CwF notation for clarity and space reasons,
we now show how the rules of \cref{sec:coverages} can be used to build
a cover:
\begin{flalign*}
        \{ & ((m + 1)\ f\ (\mathsf{cons}\ (m + 1)\ x\ (\mathsf{nil}\ \mathsf{Refl})\ \mathsf{Refl})\ \mathsf{self}),\\
&       ((m + 1)\ f\ (\mathsf{cons}\ (m + 1)\ x\ (\mathsf{cons}\ m\ y\ ys\ \mathsf{Refl})\ \mathsf{Refl})\ \mathsf{self}) \} \\ & \qquad\qquad \covers \ (n : \bN), (f : A -> A -> A),
  (v : \mathsf{Vec}\ A\ (\mathsf{suc}\ n) ), (\mathsf{self} : \ldots)
\end{flalign*}

\begin{itemize}
  \item Identity (\cref{cor:id-cover}) has variables $(n) (f) (v) (\mathsf{self})$ covering the scrutinee type;
  \item Coproduct (\cref{cor:coprod-cover}) has\! $\set{(\mathsf{nil}\,eq_{nil}), (\mathsf{cons}\,m'\,x\,xs\,eq_{cons}) } \covers \mathsf{Vec\,A\,q}$ for any $q$;
  \item Composition (\cref{thm:cover-comp})  allows us to construct the canonical cover\\
        $\set{(n\,f\,(\mathsf{nil}\,eq_{nil})\,\mathsf{self}), (n\,f\,(\mathsf{cons}\,m'\,x\,xs\,eq_{cons})\,\mathsf{self}) }$;

  \item We have $eq_{nil} : n + 1 = 0$, but this type is empty, so applying the absurd rule
        (\cref{thm:absurd-cover})
        gives a singleton cover $\set{(n\,f\,(\mathsf{cons}\,m'\,x\,xs\,eq_{cons})\,\mathsf{self})}$;
  \item $eq_{cons} : n + 1 = m' + 1$, so we get a pullback substitution mapping $m'$ to $n$ and all other variables to themselves.
        Applying the $\mathsf{Refl}$ rule (\cref{cor:refl-cover}),
        $n\,n\,\mathsf{Refl}$ is a cover of ${(m' : \bN) \cext (n : \bN) \cext (n+1 = m'+1) }$.
        By composition, the scrutinee context has singleton cover
        $\set{(n\,f\,(\mathsf{cons}\,n\,x\,xs\,\mathsf{Refl})\,\mathsf{self})}$
        in the canonical coverage;
  \item The coproduct property for $\mathsf{Vec}$ and composition yield a cover \\
        $\set{(n\,f\,(\mathsf{cons}\,n\,x\,(\mathsf{nil}\,eq_{nil})\,\mathsf{Refl})\,\mathsf{self}),
         (n\,f\,(\mathsf{cons}\,n\,x\,(\mathsf{cons}\,m\,y\,ys\,eq_{cons})\,\mathsf{Refl})\,\mathsf{self}) }$;
  \item Finally, since $eq_{nil} : n = 0$ and $eq_{cons} : n = m + 1$, we apply the $\mathsf{Refl}$ rule
        for each proof, along with composition, to obtain the desired cover above.
\end{itemize}
Then, we can use the sheaf condition model how
$f\ x\ (\mathsf{self}\ f\ (cons\ y\ ys\ \mathsf{Refl})\ \mathsf{Refl})$ and $x$
are amalgamated
into a denotation for the entire function.


\section{Discussion}
\label{sec:discussion}

\subsection{Related Work}

Dependent pattern matching was first proposed by
\citet{coquandPatternMatchingDependent1992}.
While this work contains no explicit mentions of sheaf theory,
it originated the idea that patterns could be thought of in terms of
coverings and partitions of a space, which greatly inspired our work.
The theory and practice of both pattern matching and eliminators
foundational developments in proof assistants:
\citet{mcbrideEliminationMotive2002} developed elimination for \textsc{Lego},
and later \textsc{Epigram}~\citep{mcbrideEpigramPracticalProgramming2005}.
This was extended to views and with-clauses by \citet{mcbrideViewLeft2004}.

\Citet{goguenEliminatingDependentPattern2006} show how pattern matching can
be elaborated to primitive eliminators, and hence given semantics in any model
that had semantics for eliminators.
These are in turn given semantics using
initial algebras~\cite{abbottContainersConstructingStrictly2005,altenkirchIndexedContainers2015}.
 \Citet{cockxPatternMatching2014} extend this to work with univalent theories.
 Elaboration is similar to amalgamation using the sheaf
 condition,
 but amalgamation occurs strictly in the model.
  Elaborating to eliminators also handles recursion,
which is not yet explicitly included in our sheaf semantics.


To our knowledge, the first  explicit connection between
the sheaves and pattern matching was by \citet{shermanComputableDecisionMaking2018},
which gave a framework for pattern-matching on real numbers
using topological spaces. The thesis version of this
work~\citep{shermanMakingDiscreteDecisions2017} generalizes the approach
from topological spaces to Grothendieck topologies.
\Citet{cockxOverlappingOrderIndependentPatterns2014}
give similar semantics to overlapping patterns by treating them as definitional equalities,
using confluence rather than sheaves.
Using equalizers or pullbacks to represent unification was originated by
\citet{10.5555/49340}, as well as \citet{goguen1989unification}.

\subsection{Future Work}

\subsubsection{First-Class Pattern Synonyms}
\label{subsec:discussion-patsyn}
An immediate application of this work would be to implement
an enhanced version of pattern synonyms in a language like Agda.
Currently, Agda lets the programmer declare pattern synonyms,
 but each name must map to a syntactic pattern
 i.e., a set of nested constructor applications. Agda checks if a definition is covering
 by elaborating to these
patterns.
Our framework could be used to build direct coverage checking for pattern synonyms,
so the programmer could build their own alternate, extensible covers of a type,
using sheaf theory to justify their coverage.
This would provide direct semantics for the user-defined views of \citet{wadlerViewsWayPattern1987}
and \citet{mcbrideViewLeft2004}.
Further research is needed to extract a constructive procedure
for amalgamating branches that can be implemented in practice.

\subsubsection{Overlapping Pattern Matches}

Our semantics require non-overlapping, injective patterns,
but our framework suggests a way to lift this restriction.
Recall that the sheaf condition only requires that a matching family agrees on the overlap
between covering patterns. This suggests two ways to give semantics
to overlapping patterns: by ensuring that the right-hand sides of each pattern match
agree on the overlap of their left-hand sides,
or by adding information to each pattern to ensure they are actually non-overlapping.

The latter approach matches current implementations:
catch-all patterns
are elaborated into multiple branches whose left-hand patterns are the constructors
that have not yet been used.
Unfortunately, to prove anything about a function defined this way,
the programmer needs a proof case for each branch in the elaboration, even
if they correspond to a single branch in the function as written.
Our framework may support
 canonical covers which contain extra information preventing overlap, such as proofs that previous branches had not matched.
These could be used to develop covers for matches with overlapping patterns that
do not require creating additional cases during elaboration, enabling more succinct
proofs about overlapping cases.

\subsubsection{Inductive Datatypes and Termination Checking}
\label{subsec:future-rec}

When patterns are not restricted to constructors, it is not immediately apparent
which recursive pattern matching functions can be soundly modelled,
since pattern variables may not be structurally smaller than the patterns in which they occur.
Further study is needed to devise criteria for which recursive definitions
are well founded with non-constructor patterns.


\subsubsection{Beyond Top-Level Matches}
\label{subsec:future-nest}
Our semantics only support top level pattern matches.
Many of the results we used, such as the fundamental theorem of topos theory,
are well suited to top-level matches but do not directly translate to terms
in an arbitrary context. Additionally, if the scrutinee type of a pattern match
is in a non-empty context, then matching affects not only the motive, but  may refine the values or types of variables
in the context on the left.
These technical issues  suggest a semantic theory of
\textit{telescopes}, which are objects representing extensions to a given context
by some number of types, and \textit{environments}, which extend substitutions
by some number of terms.

\subsubsection{With Clauses}
Our approach to matching on equality proofs gives an intuition for modelling
Agda-style with-clauses and views~\citep{mcbrideViewLeft2004},
though a full account is beyond the scope of this paper.
Suppose we are defining a pattern match with scrutinees of type $\D$ and result type
$\D |- T : \typeType{} $, and we want to match on some intermediate expression $\D |- s : S$.
There is an isomorphism $\iota : \D \isomorphic \D, (x : S), (pf : x =_{S} s)$.
So if we have a cover of $\setj{i}{s_{i} : \D_{i} -> \D } $ to match $s$ against, we can use composition and extension to obtain a cover
$\setj{i}{s_{i} : \D, (pf : s_{i} = s) -> \D}$.
In the case that $s_{i}$ and $s$ unify, the $\mathsf{Refl}$ rule from above
can be used to match on the equality, and in an extensional model, the goal type can
be safely rewritten due to the existence of the equality proof.

\subsubsection{Intensional Models}
\label{subsec:disc-intensional}
Our current approach relies on extensional models,
where equality proofs correspond with equality in the model.
We can define models of \lang in terms of canonical coverages and non-syntactic equality,
and we can give a CwF term model for \lang
because syntactic pattern matching fulfills the criteria of \cref{subsec:cwf-pat},
but the term model for \lang cannot be described in terms of sheaf-theoretic coverages.

To show the issue, we show that $\set{\mathsf{true},\mathsf{false}}$ is not a canonical
cover of $\mathsf{Bool}$ for the CwF given by well-typed $\lang$ terms quotiented
by definitional equality.
Consider a function
$\mathsf{haltsInN} : \mathsf{SyntaxTree} ->  \bN ->  \mathsf{Bool}$,
that looks at a syntax tree of an untyped
lambda calculus term and checks whether it halts in $n$ or fewer steps.
Consider also $\Omega : \mathsf{SyntaxTree}$, a representation of $(\lambda x \ldotp x\ x)(\lambda x \ldotp x\ x)$.
If $\set{\mathsf{true},\mathsf{false}}$ is a canonical cover,
it is also a sieve in the canonical topology~\citep[C2.1.8]{johnstoneSketchesElephantTopos2003},
so pulling the sieve back by $\mathsf{haltsInN}\ \Omega$
produces the set of arrows $\set{h_{j} \circ !_{V_{j}} : V_{j} -> \bN \mid h_{j} : \bOne -> \bN}$.
This contains each arrow in $\bOne -> \bN$,
i.e., each natural number.
For the cover to be canonical, for any type $T$ there must be a way to amalgamate
 $\set{ t_{j} : T \mid h_{j} : \bOne -> \bN}$ into  $\bN -> T$.
Then all set-theoretic infinite sequences of natural numbers
could be amalgamated into type-theoretic functions $\bN -> \bN$, which is impossible.

The above example relies on the existence of an infinite cover.
While infinite covers are allowed in sheaf theory, they do not correspond directly
to pattern matches that a programmer can write down.
So further exploration of finite and infinite covers may resolve the issue.

Another issue is that
extensional models typically imply that all equality proofs of a given
type are equal. As such our approach is incompatible with univalent theories
like Homotopy or Cubical Type Theory~\citep{univalentfoundationsprogramHomotopyTypeTheory2013,cohenCubicalTypeTheory2018}.
Both of these issues might be addressed by replacing sheaves
with stacks or, even more generally, $\infty$-stacks~\citep{lurieHigherToposTheory2009}.
These replace the strict equality of the sheaf condition with higher structure.
However, the technical and theoretical overhead of switching to stacks
is considerable, and utilizing them for pattern matching
will be a significant undertaking.

\subsubsection{Toposes and Quasi-toposes}

Apart from pattern matching, the theory of sheaves plays a central role in categorical logic,
since categories of sheaves over a coverage form \textit{Grothendieck toposes}, which serve as
models of constructive logic.
Quasi-toposes relax the sheaf conditions to only require uniqueness of amalgamations.
Future work should search for deeper connections to toposes or quasi-toposes.
Two-level type theories~\citep{annenkovTwolevelTypeTheory2023}
may yield some answers, since they describe the interactions between a model of a type theory
and the category of presheaves over that model.

\subsection{Conclusion}
This work formalizes the connection between dependent pattern matching
and the notion of sheaves over a site.
We have provided a framework which is expressive enough to capture
the semantics of current pattern matching implementations, while laying the groundwork
for future enhancements.
Our work demonstrates that elaboration to eliminators is not the only feasible
semantics for dependent pattern matching,
and that there is perspective to be gained from treating pattern matching
as a core feature and using the lens of sheaf theory.

\renewcommand{\emph}[1]{#1}

\bibliographystyle{splncs04nat}
\bibliography{refs}

\end{document}